\documentclass[floats,floatfix,showpacs,amssymb,prd,twocolumn,superscriptaddress,nofootinbib]{revtex4-1}

\newlength{\figw} 
\usepackage{graphicx,epsf, epsfig, amssymb}
\usepackage{bm}
\usepackage{longtable,tabularx}
\usepackage{color}
\usepackage[breaklinks]{hyperref}
\usepackage{amsfonts,amsmath,amssymb,mathrsfs,gensymb}
\usepackage{natbib}
\usepackage{array}
\usepackage{multirow}
\usepackage{rotating,array}
\usepackage{graphicx}

\newcommand\unit[1]{{\rm #1}}

\newcommand\optional[1]{}
\def\be{\begin{equation}}
\def\ee{\end{equation}}
\def\beq{\begin{eqnarray}}
\def\eeq{\end{eqnarray}}
\usepackage{comment}

\newcommand\citeMCMC{\cite{LIGO-CBC-S6-PE,2011PhRvD..83h2002D,2011PhRvD..84f2003C,gr-extensions-tests-Europeans2011,gwastro-mergers-PE-Aylott-LIGOATest,2011ApJ...739...99N,gwastro-mergers-HeeSuk-CompareToPE-Aligned,gw-astro-PE-lalinference-v1}}

\newcommand\apjl{ApJL}
\newcommand\mnras{MNRAS}

\newcommand\xicVal{-0.11}
\newcommand\xic{\xi_T}

\begin{document}

\title{Distinguishing black-hole spin-orbit resonances by their  gravitational wave signatures.  II:  Full parameter  estimation } 

\author{Daniele Trifir\`o}
\email{daniele.trifiro@ligo.org}
\affiliation{Dipartimento di Fisica E. Fermi, Universit\`a di Pisa, Pisa 56127, Italy}
\affiliation{Department of Physics and Astronomy, The University of
Mississippi, University, MS 38677, USA}

\author{Richard O'Shaughnessy}
\email{rossma@rit.edu}
\affiliation{Center for Computational Relativity and Gravitation, Rochester Institute of Technology, Rochester, NY 14623, USA}

\author{Davide Gerosa}
\email{d.gerosa@damtp.cam.ac.uk}
\affiliation{Department of Applied Mathematics and Theoretical Physics, Centre for Mathematical Sciences, University of Cambridge, Wilberforce Road, Cambridge CB3 0WA, UK}

\author{Emanuele Berti}
\email{eberti@olemiss.edu}
\affiliation{Department of Physics and Astronomy, The University of
Mississippi, University, MS 38677, USA}
\affiliation{CENTRA, Departamento de F\'isica, Instituto Superior T\'ecnico, Universidade de Lisboa, Avenida Rovisco Pais 1, 1049 Lisboa, Portugal}

\author{Michael Kesden}
\email{kesden@utdallas.edu}
\affiliation{Department of Physics, The University of Texas at Dallas, Richardson, TX 75080, USA }

\author{Tyson Littenberg}
\email{tyson.littenberg@northwestern.edu}
\affiliation{Center for Interdisciplinary Exploration and Research in Astrophysics (CIERA) \& Department of Physics and Astronomy, Northwestern University, 2145 Sheridan Road, Evanston, IL 60208}

\author{Ulrich Sperhake}
\email{u.sperhake@damtp.cam.ac.uk}
\affiliation{Department of Applied Mathematics and Theoretical Physics, Centre for Mathematical Sciences, University of Cambridge, Wilberforce Road, Cambridge CB3 0WA, UK}
\affiliation{Department of Physics and Astronomy, The University of
Mississippi, University, MS 38677, USA}
\affiliation{California Institute of Technology, Pasadena, CA 91125, USA}


\date{\today}

\begin{abstract}
  Gravitational waves from coalescing binary black holes encode the evolution of their spins prior to merger.
  In the post-Newtonian regime and on the precession timescale, this evolution  has one  of three
   morphologies, with the spins  either librating around one of two fixed points (``resonances'') or circulating
  freely. 
  In this work we perform full parameter estimation on resonant binaries
  with fixed masses and spin magnitudes, changing three parameters: a conserved ``projected effective
  spin'' $\xi$ and resonant family $\Delta\Phi=0,\pi$ (which uniquely label the source); the inclination
  $\theta_{JN}$ of the binary's total angular momentum with respect to
  the line of sight (which determines the strength of precessional
  effects in the waveform); and the signal amplitude.  We demonstrate that resonances can be distinguished for a wide range of binaries,
  except for highly symmetric configurations where precessional
  effects are suppressed.  Motivated by new insight into
  double-spin evolution, %
  we introduce new variables to characterize precessing black hole binaries
  which naturally reflects the timescale separation of the system and therefore
  better encode the dynamical information carried by gravitational waves.%
\end{abstract}
\maketitle

\section{Introduction}

Gravitational waves (GWs) from binary black holes (BBHs) are expected
to be an important source \cite{2010CQGra..27q3001A,2012ApJ...759...52D,2013ApJ...779...72D,2015ApJ...806..263D} for future
networks of GW detectors such as Advanced LIGO/Virgo
\cite{0264-9381-32-7-074001,0264-9381-32-2-024001}, LIGO-India \cite{2013IJMPD..2241010U},
KAGRA \cite{2012CQGra..29l4007S,0264-9381-31-22-224003}, and the Einstein Telescope
\cite{2010CQGra..27s4002P}.  Once formed, BBHs emit GWs that extract
energy and angular momentum from the orbit, decreasing the binary
separation and increasing the orbital frequency (and thus the GW
frequency). Most binaries are expected to circularize by the time they
enter the sensitivity band of ground-based detectors
\cite{1963PhRv..131..435P,1964PhRv..136.1224P} (see
e.g. \cite{Sperhake:2007gu,Hinder:2008kv,2009PhRvD..80h4001Y,2013PhRvD..87d3004E,2014PhRvD..90h4016H}
and references therein for recent work on eccentric binary waveforms,
rates and detection strategies).
Circular BBH inspirals are characterized by the masses ($m_1,\,m_2$)
and spins ($\mathbf S_1,\,\mathbf S_2$) of each black hole.
The spin and orbital angular momenta precess on timescales shorter
than the radiation-reaction timescale
\cite{1994PhRvD..49.6274A,1995PhRvD..52..821K,BCV:PTF}.  Their general
evolution can be understood relative to the two one-parameter families
of fixed points (``post-Newtonian resonances'') of the orbit-averaged
spin precession equations \cite{2004PhRvD..70l4020S}. At resonance,
the spins and orbital angular momentum $\mathbf{L}$ remain
coplanar. If $\Delta \Phi$ denotes the angle between the projection of
the spins in the orbital plane orthogonal to $\mathbf{L}$, the spins
can either lie on the same side of $\mathbf{L}$ (the $\Delta\Phi=0$
resonance) or on opposite sides of $\mathbf{L}$ (the $\Delta\Phi=\pi$
resonance).  In general, the conservative evolution of the spins of
any BBH can be understood as either a trapped phase-space orbit around
one of the two fixed points ($\Delta\Phi=0$ and $\Delta\Phi=\pi$) or
as an orbit where $\Delta\Phi$ circulates freely. Therefore the phase
space can be split into three classes, or ``morphologies''
\cite{2015PhRvL.114h1103K,2015PhRvD..92f4016G}.
The morphology of BBHs in the
Advanced LIGO/Virgo band is determined by the spin configuration when the compact binary is first formed, long prior to its
detection via GWs  \cite{2013PhRvD..87j4028G}.
This relationship may enable GW measurements to constrain the efficiency of tidal interactions  in binary star evolution and may indicate whether a mechanism such as mass transfer, stellar
winds, or supernovae can induce mass-ratio reversal (so that the
heavier black hole is produced by the initially lighter stellar
progenitor).  More broadly, spin measurements of BBHs have long been expected to be a critical element in
distinguishing the underlying astrophysical model for compact binary formation, complementing other
measurements: see e.g.~\cite{2008ApJ...682..474B,2010CQGra..27k4007M,2014PhRvD..89j2005O,2015ApJ...810...58S}.
Gravitational radiation encodes the dynamics and properties of the
inspiralling binary. Precessing binaries produce a rich, highly
modulated signal
\cite{1994PhRvD..49.6274A,1995PhRvD..52..821K,BCV:PTF,gwastro-mergers-nr-Alignment-ROS-PN,gwastro-mergers-nr-Alignment-ROS-Polarization,gw-astro-SpinAlignedLundgren-FragmentA-Theory,gwastro-SpinTaylorF2-2013,2013PhRvD..88l4015K,Mroue:2013xna},
enabling constraints on compact binary parameters, such
as masses and the misaligned spins
\cite{2008ApJ...688L..61V,2008CQGra..25r4011V,gwastro-mergers-HeeSuk-FisherMatrixWithAmplitudeCorrections,gwastro-mergers-HeeSuk-CompareToPE-Precessing,gwastro-pe-systemframe,2014PhRvL.112y1101V,LIGO-CBC-S6-PE}.
This information can be recovered from detector measurements by
systematically comparing all possible candidate signals with the data
and constructing a Bayesian posterior probability distribution for all
binary parameters \citeMCMC{}.
Out of all precessing BBH configurations possible a priori, the two librating morphologies are relatively common for comparable-mass
binaries in the close orbits to which LIGO is sensitive, but relatively rare for binaries drawn from a uniform mass
distribution \cite{2015PhRvD..92f4016G}. %
That said, because the relative spin dynamics do not directly source
the dominant GW signal, it is important to investigate the extent to
which the relative orientation of precessing spins can be constrained in general, and
the degree to which measurements distinguish morphologies in
particular. Vitale et al. \cite{2014PhRvL.112y1101V} recently studied
this problem for a small sample of precessing binaries, claiming that
the relative angle $\Delta\Phi$ between the projection of the two
spins into the orbital plane cannot be measured.
Previously, Gerosa et al.~\cite{2014PhRvD..89l4025G} examined the GW
signal from the two resonant families and evaluated the overlap between
the waveforms with all parameters fixed, to demonstrate that the two families
produce qualitatively and quantitatively distinct GW signatures.  
\citet{2014CQGra..31j5017G} used overlaps to imply  resonant and non-resonant binaries were distinguishable.  
In
this work we apply the
state-of-the-art \textsc{lalinference} parameter estimation code
\cite{gw-astro-PE-lalinference-v1} to the expected detector response
due to resonant post-Newtonian binaries in the nearby Universe.
Using the posterior distribution of compact binary parameters, we
assess for the first time how confidently these measurements
distinguish between the three possible morphologies.  As we discuss in
detail, morphological classification is related to (but not strictly
dependent on) the accuracy with which we can measure the angle
$\Delta \Phi$ between the projection of the two spins on the orbital
plane.  On physical grounds, morphological classification is far more robust: while $\Delta\Phi$ changes on the
precession timescale, a binary's 
morphology only changes on the inspiral timescale. 

The paper is organized as follows. In Section~\ref{sec:Methods} we
briefly review GW parameter estimation, the specific grid of binaries
examined for this work and the parameters needed to interpret the
posterior distribution in the context of post-Newtonian resonances.
In Section~\ref{sec:Results} we show that the morphology of exactly
resonant binaries can be reliably determined at astrophysically
plausible signal amplitudes, unless the binary is face-on.  Motivated
by a recent analytic solution of the spin precession equations
\cite{2015PhRvL.114h1103K,2015PhRvD..92f4016G}, we introduce two
coordinate systems to more naturally characterize how well properties
of double-spin binaries have been constrained.  
We conclude in Section~\ref{sec:Conclusions} describing how our
results impact the broader program of astrophysical inference using GW
measurements.  In an attempt to keep this paper self-contained, we
report various technical material in the
appendices. Appendix~\ref{ap:coord} describes the relationship between
several conventions used to describe precessing spins that have
appeared in the literature.  Appendix~\ref{ap:alpha} reviews the
characteristic precession timescales introduced in our previous work
on analytic solution of the two-spin precession equations
\cite{2015PhRvL.114h1103K,2015PhRvD..92f4016G}.  Finally, in
Appendix~\ref{ap:details} we provide some technical details on our
parameter estimation procedure.

\section{\label{sec:Methods}Methods}

\subsection{\label{subsec:candidatesources}Candidate sources and signals}

\begin{figure}
\includegraphics[width=\columnwidth]{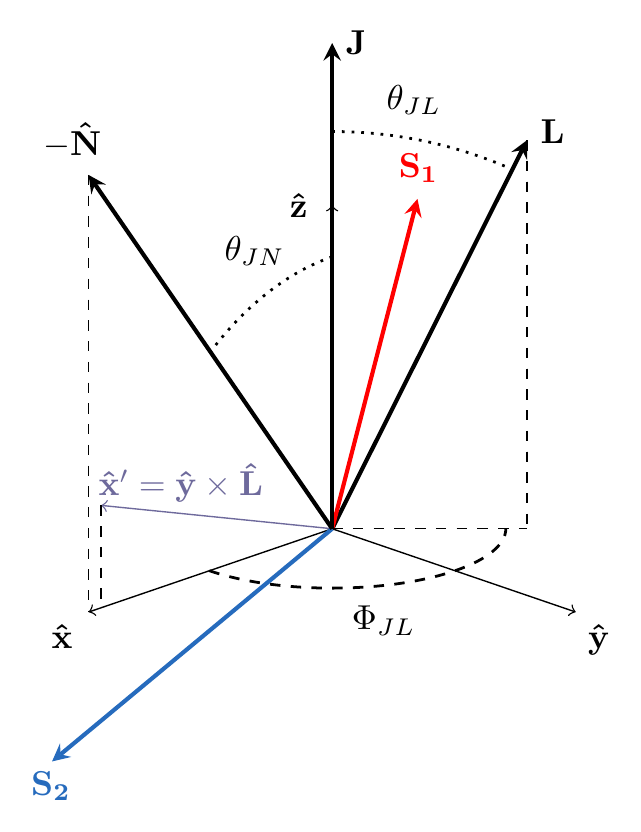}
\includegraphics[width=\columnwidth]{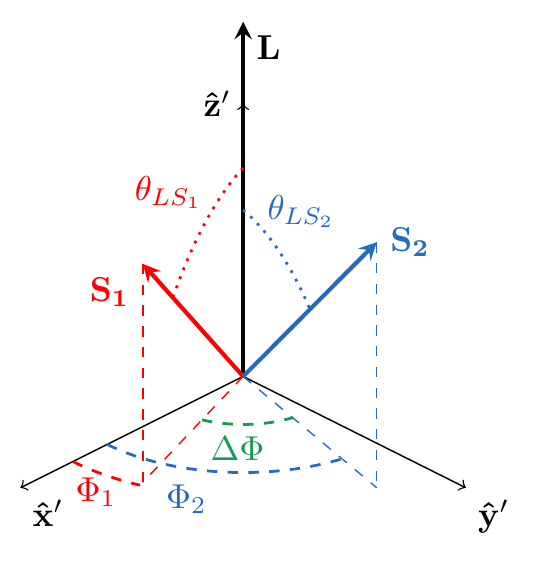}
\caption{\label{fig:coordinates} Definitions of the angles used throughout
  this paper. The directions of the spins ${\bf S}_{1}$ and
  ${\bf S}_{2}$ are specified using polar angles
  ${\bf \theta}_{LS_{i}}$ and azimuthal angles $\Phi_{i}$
  ($i = 1, 2$), measured relative to ${\bf \hat{x}^\prime}={\bf \hat{x}} \times {\bf \hat{L}}$.
  For resonant binaries, the orbital angular momentum
  and spins remain coplanar, implying that the angle
  $\Delta\Phi = \Phi_{2}-\Phi_{1}$ remains constant: either
  $\Delta\Phi = 0$ or $\Delta\Phi = \pi$.  
}
\end{figure}

The GW signal from the binary depends on $8$ intrinsic (physical)
parameters -- the component masses $m_{1}$ and $m_{2}$ and spin
vectors ${\bf S}_{1}$, ${\bf S}_{2}$ ($3$ components each) -- and $7$
extrinsic parameters -- the event time $t_{\rm ref}$; luminosity
distance $D_{L}$; sky location or propagation direction relative to
the Earth's equatorial coordinate system, expressed via two angles
$\alpha,\delta$ (right ascension and declination);
and three Euler angles describing the orientation of the binary
relative to the Earth.  The rotation relating the Earth frame and the
binary's frame is commonly characterized using the line of sight
$\mathbf{\hat{N}}$ from the observer to the binary and the
time-dependent orbital angular momentum direction $\mathbf{\hat{L}}$,
evaluated at some fiducial orbital frequency.  For example, the
orbital orientation is often characterized by the inclination angle
$\iota$, defined by $\cos\iota= -{\bf \hat{L}}\cdot{\bf \hat{N}}$ (in
this convention, a binary whose angular momentum points towards the
observer corresponds to $\iota=0$.)
Among different possible parametrizations for these intrinsic and extrinsic parameters, we choose to use the following \cite{gwastro-pe-systemframe}:
\begin{equation}\label{eq:parametrization}
	{\boldsymbol\theta} = \left\{m_{1}, m_{2}, \chi_{1}, \chi_{2}, \theta_{JN}, \theta_{LS_{1}}, \theta_{LS_{2}}, \Delta\Phi, \phi_{JL}\right\},
\end{equation}
as well as coalescence event's 4 spacetime
coordinates,  polarization, and orbital phase, which will be marginalized over and not discussed henceforth. In this list, 
the symbol $m_{i}$ denotes the component masses, $\chi_{i}={|{\bf S_{i}}|}/{m_{i}^{2}}$ 
the adimensional spin magnitudes, $\theta_{L
  S_{i}}$ 
 the angle between the orbital angular momentum ${\bf L}$ and
the spin vector ${\bf S}_{i}$, $\theta_{JN}$ the angle
between the total angular momentum ${\bf J} = {\bf L} + {\bf S}_{1} + {\bf S}_{2}$ and the line of sight to the observer
$(-{\bf
  \hat{N}})$, $\Delta\Phi$ the azimuthal angle between ${\bf S}_{1}$ and ${\bf S}_{2}$, and $\phi_{JL}$ the
azimuthal angle between ${\bf J}$ and ${\bf L}$, relative to a reference phase defined via the projection of $-\hat{{\bf N}}$
onto the orbital plane; see Figure \ref{fig:coordinates} and \cite{gwastro-pe-systemframe}. %
All parameters are specified when the GW frequency (twice the orbital frequency) is $f_{\rm ref}=100\unit{Hz}$.   Except
for $m_1,m_2,\chi_1,\chi_2$, all of our intrinsic parameters evolve during the inspiral.  %

The orbital dynamics and GW signal are constructed via the 
\textsc{lalsimulation} SpinTaylorT4 code \cite{lal}.  
Based on previous implementations \cite{2003PhRvD..67j4025B,2004PhRvD..70j4003B},  this time-domain code solves the orbital dynamics of
an adiabatic, quasicircular inspiraling binary using the ``TaylorT4'' method
\cite{gw-astro-PN-Comparison-AlessandraSathya2009}  for the phase evolution and
(orbit-averaged) precession equations for the angular momenta \cite{1995PhRvD..52..821K}.   
In all cases we used a GW phase up to 3.5PN in
nonspinning terms;  up to 2.5PN in spin-orbit terms; up to 2PN spin-spin terms;
and orbit-averaged spin precession equations up to 2PN, including the
leading-order spin-orbit and spin-spin interactions.  
As described below, we performed calculations both with and
without  black
hole-consistent coefficients in composition-dependent terms like the
quadrupole-monopole (QM)
coupling. We evaluate the GW amplitude $h(t)$ using only the
leading-order (Newtonian) quadrupole \cite{WillWiseman:1996}.

To compare waveforms from binaries belonging to different morphologies
in controlled conditions we investigate only a specific combination of
the binaries' intrinsic parameters: total mass $M=13.5 M_{\odot}$,
mass ratio $q={m_{2}}/{m_{1}}=0.8$ (which implies
$m_{2}=6.0 M_{\odot}$ and $m_{1}=7.5M_{\odot}$), and adimensional spin
magnitudes $\chi_{1}=\chi_{2}=1$ (maximally spinning BHs). Our choice, motivated by population-synthesis
predictions~\cite{2012ApJ...759...52D,2013ApJ...779...72D,2015ApJ...806..263D} and computational restrictions\footnote{Data
 presented in this paper involves full MCMC simulations which required several million CPU hours over the course of several months to compute.}, also facilitates comparisons with our previous works~\cite{2013PhRvD..87j4028G,2014PhRvD..89l4025G}. 

Following \cite{2014PhRvD..89l4025G}, we parametrize binaries from each morphology using the dimensionless projected effective spin 
 $\xi$:
\begin{equation}
\label{eq:xi}
	\xi \equiv \frac{{\bf S}_{0} \cdot {\bf \hat{L}}}{M^{2}}\Big|_{f=f_{\rm ref}} = \frac{\chi_{1}\cos\theta_{LS_{1}} + q\ \! \chi_{2}\cos\theta_{LS_{2}}}{1+q}
\end{equation}
where ${\bf S}_{0}$ is the effective spin
\begin{equation}
	{\bf S}_{0} \equiv 	\left(1+q\right) {\bf S}_{1} + \left(1+\frac{1}{q}\right) {\bf S}_{2} \; .
\end{equation}
The projected effective spin $\xi$, which was first introduced in
\cite{2001PhRvD..64l4013D,2008PhRvD..78d4021R} and used to solve the
spin precession equations
in~\cite{2015PhRvL.114h1103K,2015PhRvD..92f4016G}, is a constant of
motion at 2PN order in spin precession and 2.5PN order in radiation
reaction.

\begin{figure*}
\includegraphics[width=2.0\columnwidth]{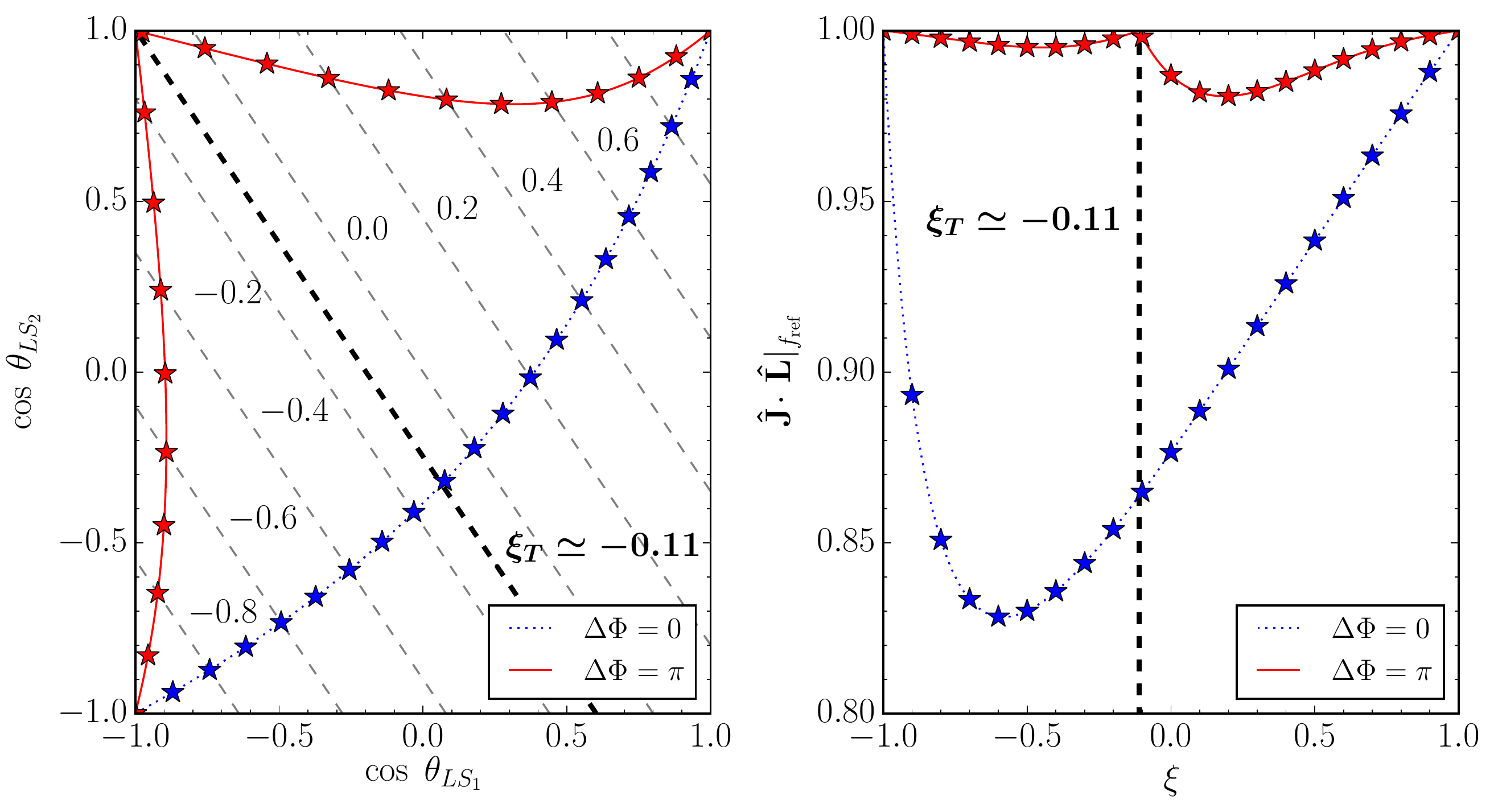}
\caption{\label{fig:Recap}\textbf{Review of resonant dynamics and our
    simulations.}  \emph{Left panel}: Illustration of the specific two
  one-parameter families of resonant binaries used in this work.  The
  two curved paths show $\cos\theta_{LS_1},\cos\theta_{LS_2}$ for
  $\Delta\Phi=0$ (blue and dotted) and $\Delta\Phi=\pi$ (red and solid), superimposed on
  contours of constant $\xi$ (dashed). $\xi=\xic\simeq\xicVal$ is shown with a thick dashed line.
  Stars mark the location of each
  of our simulations (blue for $\Delta\Phi=0$, red for
  $\Delta\Phi=\pi$).  \emph{Right panel}: Plot of
  $\hat{\mathbf{J}}\cdot \hat{\mathbf{L}}$ versus $\xi$ for each
  resonant family at the reference frequency $f_{\rm ref}=100~\unit{Hz}$.  As
  discussed in the text, our source binaries in the $\Delta\Phi=\pi$
  resonance have spins and $\mathbf{L}$ much more closely aligned with
  each other than sources in the $\Delta\Phi=0$ resonance. All source binaries have  $L>S_1+S_2$, and hence dynamics characterized by the
two-spin solution described in \cite{2015PhRvL.114h1103K, 2015PhRvD..92f4016G}.
}
\end{figure*}

Our candidate sources will be exactly resonant, which is a conservative choice, since in this case the waveforms are
weakly modulated and our ability to constrain parameters correlates with the modulation in the waveforms.
For resonant binaries, for each $\xi$ and each family there are unique values of
$\theta_{LS_1},\theta_{LS_2},\Delta\Phi$, meaning that $(\xi,\Delta\Phi)$ uniquely label the source.
For simplicity, we fix $\phi_{JL}=0$ at $f_{\rm ref}=100\unit{Hz}$, so
$\mathbf{\hat{N}},\mathbf{L},\mathbf{J}$ are coplanar.  
Finally, because the amount of spin-induced precession seen by the observer in the GW signal depends on $\theta_{JN}$
\cite{2014PhRvD..89l4025G}, we explore a range of $\theta_{JN}$ between $0$ and $\pi/2$ without loss of generality: the posterior distribution for
sources with $\theta_{JN}>\pi/2$ can be related to a posterior with $\theta_{JN}<\pi/2$ by suitable reflection symmetry. 
Hence, having chosen the resonant family ($\Delta\Phi=0$ or $\Delta\Phi=\pi$);  the inclination angle $\theta_{JN}$ ($0$,
${\pi}/{8}$, ${\pi}/{4}$, ${3}\pi/{8}$, ${\pi}/{2}$); and  $\xi$ in the $[-1,1]$ interval in steps
of $0.1$, we obtain 42 candidate sources and 210 candidate signals.
For simplicity we adopt the sky location  $(\alpha, \delta) =
(0\degree,0\degree)$ and fiducial GPS (Global Positioning System) time equal to zero.
To better
isolate the intrinsic differences between these signals, we require
all sources to produce the same 
coherent amplitude in a network consisting of
the three (two Advanced LIGO plus Virgo) interferometers, using the design
sensitivity curves \cite{LIGO-aLIGODesign-Sensitivity}.  Specifically,
we select the distance for each source such that the network
signal-to-noise ratio (SNR) is 20, a high but not unreasonable value
for first detections
\cite{2010CQGra..27q3001A,lrr-2016-1,LIGO-2014-WhitePaper-Data}.
This procedure defines two families of candidate sources, one for
$\Delta\Phi=0$ and one for $\Delta\Phi=\pi$, each of which is
characterized by two parameters ($\xi$, defined in Eq. \ref{eq:xi}, and $\theta_{JN}$), and produces a
well-defined strain response in each idealized instrument.  For all
masses and spins, the family with $\Delta\Phi=\pi$ has a special point
$\xi=\xic\equiv (\chi_1- q \chi_2) / (1+q)$  such that both spins are parallel to the orbital angular
momentum and antialigned with respect to one another, i.e.
$(\theta_{LS_1},\theta_{LS_2})=(\pi,0)$ \cite{2015PhRvL.115n1102G}. For our choice of binary
parameters, $\xic\simeq \xicVal$ [see Eq.~(\ref{eq:xi}) and
Figure~\ref{fig:Recap}].

Rather than explore an ensemble of noise realizations, following
previous
studies~\cite{gwastro-mergers-HeeSuk-CompareToPE-Aligned,gwastro-mergers-HeeSuk-CompareToPE-Precessing,2015ApJ...807L..24L}
we adopt a unique preferred noise realization (exactly zero noise) for
all simulations and all instruments; see Appendix \ref{ap:details} for
details.  As discussed below, we have repeated parts of the analysis
using randomly selected detector noise to demonstrate that our results
are robust with respect to this choice.
Simulation properties are summarized briefly in Table \ref{tab:sims};
in the Appendix, Table \ref{tab:fend} provides an explicit list,
including derived parameters (such as $J$ and the termination
frequency).

\begin{table}[t]
\begin{tabular}{l|r}
Parameter & Value(s) \\\hline
$m_1$ &$ 7.5~M_\odot$ \\
$m_2$ & $6~M_\odot $\\
$\chi_1$ & $1.0$\\
$\chi_2$ & $1.0$\\ 
$L/M^2$ & 0.896 \\
$\Delta\Phi$ & 0, $\pi$ \\
$\xi$ & -1, -0.9, $\ldots $, 1 \\
$\theta_{JN}$ & $0,\frac{\pi}{8},\frac{\pi}{4},\frac{3\pi}{8},\frac{\pi}{2}$ 
\end{tabular}
\caption{\label{tab:sims}\textbf{Overview of simulated binaries.} Input binary parameters at $f=100\unit{Hz}$ (equivalent to $v=\sqrt{\pi
    f{\cal M}}\simeq 0.208$, where
    $\mathcal{M}\equiv (m_1m_2)^{3/5}/(m_1+m_2)^{1/5}$ is the chirp mass)
for the  $2\times 21\times 5 $ distinct parameter
  estimation calculations described in the text.  %
}
\end{table}

As reviewed at length in~\cite{2014PhRvD..89l4025G}, each
one-parameter family of resonances has distinctly different angular
momenta orientations as a function of $\xi$: in other words, varying
$\xi$ significantly changes the dynamics of the source binary.  For
example, as shown in the left panel of Figure \ref{fig:Recap}, our
one-parameter family of dynamically unique binaries with
$\Delta\Phi=0$ has both spins comparably
($\cos \theta_{LS_1}\simeq \cos \theta_{LS_2}$) and often
significantly misaligned with $\mathbf{L}$.  As a result, the
$\Delta\Phi=0$ binaries have significant misalignment between
$\mathbf{L}$ and $\mathbf{J}$, as seen in the right panel of Figure
\ref{fig:Recap}.  By contrast, our family of $\Delta\Phi=\pi$ sources
all have either one or the other spin nearly aligned with the orbital
angular momentum: $\cos \theta_{LS_2} \simeq 1$ for $\xi> \xic$ and
$\cos \theta_{LS_1} \simeq -1$ for $\xi<\xic$.  As shown in the right
panel of Fig.~\ref{fig:Recap}, our $\Delta\Phi=\pi$ binaries have
minimal misalignment between $\mathbf{L}$ and $\mathbf{J}$,
particularly when $\xi<\xic$.
We will see below that some measurements are strongly correlated with
the amplitude of precession-induced modulations, and therefore
correlate strongly with the misalignment angle
$\theta_{JL}(\xi)=\arccos \hat{\mathbf{J}}\cdot \hat{\mathbf{L}}$.
Note also that, for our simulations, the values $\xi=\pm 1$ correspond to
spin-aligned, non-precessing binaries.

\subsubsection*{The quadrupole-monopole term}

Consistent with all previous parameter estimation studies  and the
implementation of \textsc{lalsimulation} at the time this project
began, our spin precession equations explicitly omitted the QM
coupling described in \cite{2008PhRvD..78d4021R}.
After realizing the omission, we have added the missing terms to the
\textsc{lalsimulation} code, and we have repeated parts of our
computationally expensive analysis including the QM interaction.  The conclusions of this
exercise are encouraging, and our main results are very robust under
perturbations: as discussed in Sec.~\ref{sec:systematics} below, the
posterior distributions obtained with and without the QM interaction
are basically indistinguishable, at least for near-resonant binaries.

The omission of the QM term has several important consequences from a
theoretical point of view: (i) The projected effective spin $\xi$ is
only approximately conserved in our simulated binaries. If $\xi$ is
not conserved, spin precession in general will not be
quasi-periodic. Fortunately, as argued in Sec.~\ref{sec:systematics}
below, $\xi$ is nearly conserved even in the absence of the QM
interaction for near-resonant binaries, so this does not have a
dramatic impact on the present analysis, but the omission of the QM
interaction could have larger effects for binaries that are far from
resonance.  To eliminate any ambiguity, $\xi$ (like all spin-dependent
intrinsic parameters) must be evaluated at some reference frequency,
here chosen to be $100 \unit{Hz}$. (ii) The morphological classification described
in~\cite{2015PhRvL.114h1103K,2015PhRvD..92f4016G} is no longer exact:
for example, binaries that should be exactly resonant can, in
practice, perform small librations around resonant configurations, and
some binaries may not belong to any of the three morphologies
discussed in~\cite{2015PhRvL.114h1103K,2015PhRvD..92f4016G}.
(iii) Precessional dynamics is obviously modified at 2PN, and so is the
evolution of the binary on the radiation-reaction timescale.

\begin{figure*}
\begin{tabular}{cc}
\hspace{-.5cm}\includegraphics[width=\columnwidth]{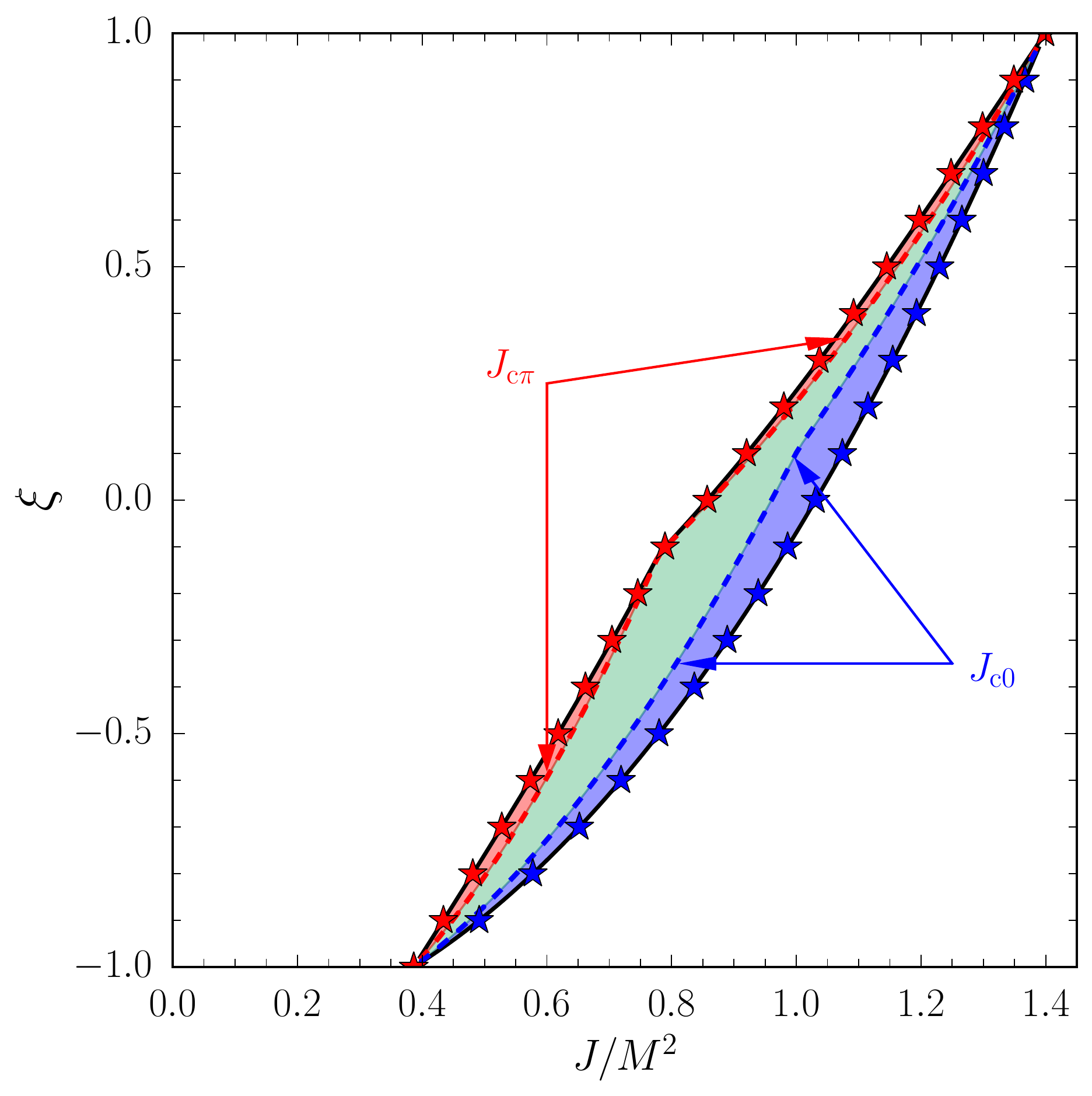}
&
\includegraphics[width=\columnwidth]{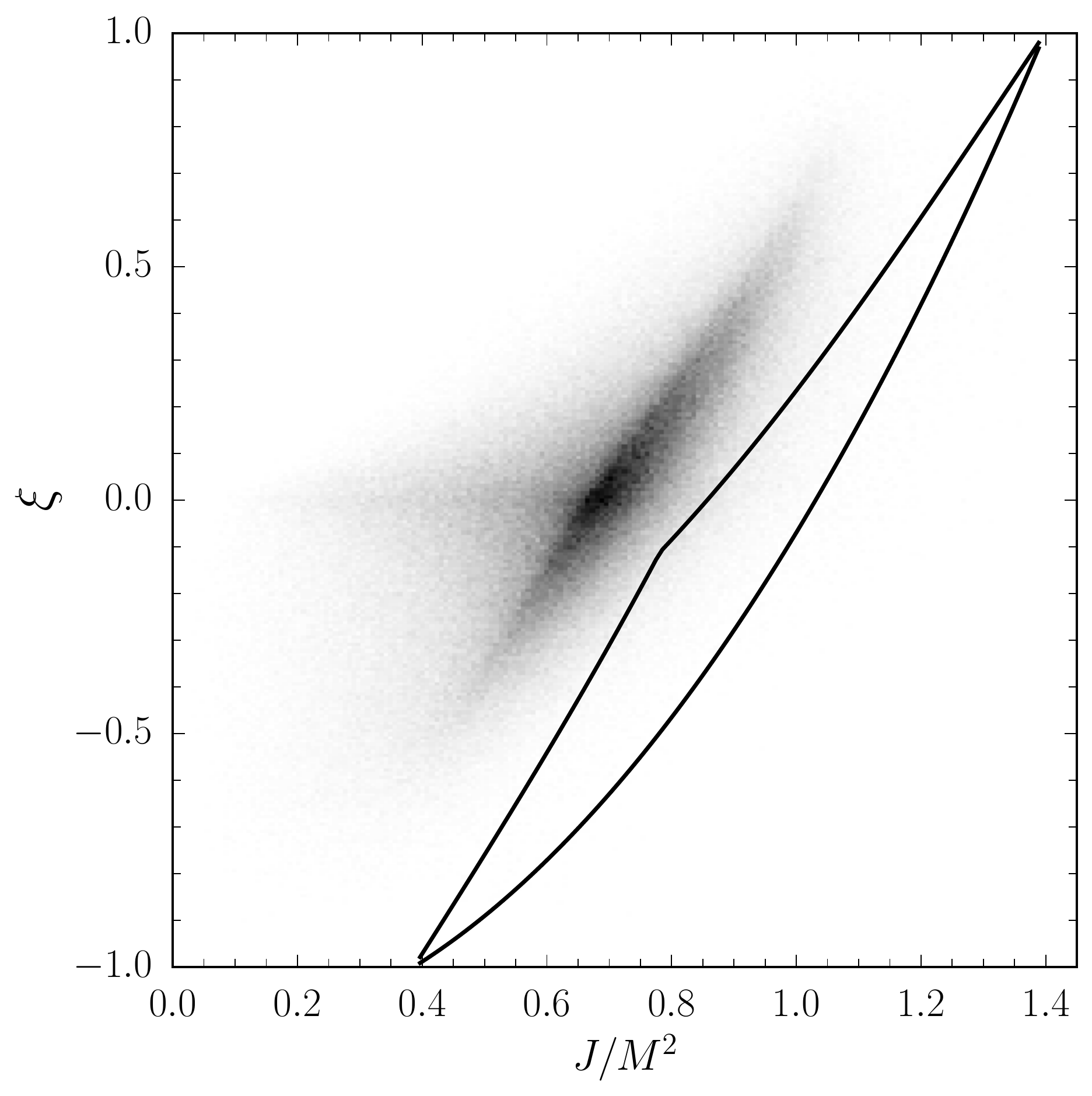}\\
\end{tabular}
\caption{\label{fig:BoundaryJXi}\textbf{The $J\xi$-plane.} For fixed
  source parameters $m_1,m_2,\chi_1,\chi_2$ and $f_{\rm ref}$, the
  solid black lines indicate the extent of the allowed values of $J$
  and $\xi$.  Any point on the lower (upper) edge of
  this region corresponds to a $\Delta\Phi=0$ ($\Delta\Phi=\pi$)
  post-Newtonian resonance.  The specific parameter configurations
  used in this paper are illustrated by red  stars (on the upper boundary)  and blue stars (on the lower boundary), as in
  Figure \ref{fig:Recap}.  
  \emph{Left:} For these same fixed source parameters, three shaded regions (red on top; blue on the bottom; and
  green in between) show the regions for, respectively, librating about
  $\Delta\Phi=\pi$, about $\Delta\Phi=0$, and circulating precession
  morphologies (described in the text). Dashed lines separating these regions (marked by arrows) show
  morphology boundaries $J_{c0,c\pi}(\xi)$.  \emph{Right:} Prior 
  probability $p(J,\xi)$,  derived from the priors described in
  \ref{sec:PEbasics}  by marginalizing out    $m_1,m_2,\chi_1,\chi_2$.
  For scale, the black curve also shows the
  same boundary of allowed $(J,\xi)$ values shown in the left panel.
The structure in this distribution reflects the broad range of mass ratios and spin orientations included in the prior; see the text for details.}
\end{figure*}

\subsection{Gravitational wave parameter estimation for double-spin binaries}\label{sec:PEbasics}

Parameter estimation can be performed using Bayes' theorem: a prior
probability distribution $p(\theta|{\cal H})$ (where ${\cal H}$ is the
hypothesis and $\theta$ represents collectively the parameters of the
system) is updated upon receiving data $d$ from the experiment to give
a posterior distribution $p(\theta|d, {\cal H})$
\begin{equation}
\label{eq:Bayes}
	p(\theta|d, {\cal H}) = \frac{p(\theta|{\cal H})p(d|\theta,{\cal H})}{p(d|{\cal H})},
\end{equation}
where $p(d|\theta,{\cal H})$ is called the ``likelihood''.  Starting
from an appropriate prior distribution, samples of the posterior
distribution are randomly selected through a stochastic sampler using
information from the data. We perform parameter estimation on
simulated GW signals from resonant binaries using the \textsc{
  lalinference} suite \cite{gw-astro-PE-lalinference-v1}, in the \textsc{
  lalinference\_mcmc} implementation.  This Markov Chain Monte Carlo
(MCMC) algorithm generates a series of independent samples from the
posterior distribution, given prior probabilities for all physical
parameters.
Consistent with previous work \cite{gw-astro-PE-lalinference-v1}, we a priori assume component masses are uniformly
distributed between $1  M_\odot$ and $30 M_\odot$, with total mass less than $35 M_\odot$;
component spins randomly oriented, with uniformly distributed
amplitude; and the merger event occurs within $\pm 0.1\unit{s}$ of our
proposed event time $t_{\rm ref}$.  To account for the high mass of
our candidate sources and the high sensitivity of the advanced
instruments used in this study, we allow the physical position to be
uniformly distributed on a sphere of Euclidean radius $6 \unit{Gpc}$,
implying uniformity over the sky and in distance -- i.e.,
$p(d)\propto d^2$ -- out to $6 \unit{Gpc}$.  The prior was not modified
to model selection bias.
This prior includes extremely high
mass ratios and a wide range of spins, favoring binaries dominated by a single spin rather than systems showing strong
two-spin effects; our conclusions drawn using this unfavorable and astrophysically unmotivated prior  should therefore
be taken as conservative underestimates.
For simplicity and consistent with previous work, cosmological effects
are not included in our calculations.\footnote{The most distant
  sources in our sample, with $\theta_{JN}\simeq 0$ and
  $\xi\simeq \pm 1$, were placed at $d_L\simeq 815\unit{Mpc}$, or
  $z\simeq 0.2$ with current cosmological parameters; typical sources
  have $d_L\simeq 500\unit{Mpc}$, or $z\simeq 0.12$. Because the GW
  strain depends only on the redshifted masses, the posterior
  including cosmological effects can be derived from a cosmology-free
  posterior by a coordinate transformation, using a modified
  (Euclidean) distance prior rather than one consistent with
  cosmology.  At the distances of interest, a Euclidean and
  cosmological distance prior nearly agree.  }
Finally, we stop our calculations when the final Markov Chain, after downsampling to remove correlations, has roughly 2,000
uncorrelated samples from the posterior \cite{gw-astro-PE-lalinference-v1}.
The parameters of the posterior samples can be represented in any
coordinate system, allowing us to efficiently compute the posterior
$p(\theta)$ using \emph{any} parameters $\theta$, independent of the
coordinates used internally by the MCMC simulation. The relative orientation of the
two spins with respect to the orbital angular momentum, $\theta_{LS_{1}}$ and $\theta_{LS_{2}}$,
can be used to characterize the system. Motivated by the
analysis of \cite{2015PhRvL.114h1103K,2015PhRvD..92f4016G} and as
described at greater length in Appendix \ref{ap:coord}, we replace these angular variables with
$(J,\xi)$, where $\xi$ is given by Eq.~(\ref{eq:xi}) and 
\begin{align}
&J=\left[ L^{2} + S_{1}^{2}+S_{2}^{2} + LS_{1} \cos \theta_{LS_{1}} + LS_{2} \cos \theta_{LS_{2}}  \notag\right.\\
&+2S_{1}S_{2}\left(\sin\theta_{LS_{1}}\sin\theta_{LS_{2}}\cos\Delta\Phi + \cos\theta_{LS_{1}}\cos\theta_{LS_{2}}\right) \left.\right]^{\frac{1}{2}}
\end{align}
is the magnitude of the total angular momentum, $J = |\mathbf{J}|$. For each resonant family ($\Delta\Phi=0,\pi$), each pair $(J,\xi)$ uniquely specifies
the binary configuration (see Fig.~\ref{fig:BoundaryJXi}).
The magnitude $L\equiv |\mathbf{L}|$ of the orbital angular momentum
is calculated at leading (Newtonian) order:
\mbox{$L=\eta M^2/v = \eta M^2/(\pi f_{\rm ref} M)^{1/3}$}.
Unlike the system-frame parameters described above, these parameters
naturally reflect the separation of timescales in the two-spin
problem, with $\xi$ conserved up to 2PN order on all timescales by the orbit-averaged spin-precession equations;
$J$ changing on the radiation-reaction timescale; and $\Delta\Phi$ changing on the
precession timescale \cite{2015PhRvL.114h1103K,2015PhRvD..92f4016G}.
For each fixed $m_1,m_2,\chi_1,\chi_2$ and $f_{\rm ref}$, a range of $(J,\xi)$ %
is allowed; see~\citet{2015PhRvD..92f4016G} for details.  %
For single-spin binaries, the relationship between $J$ and $\xi$ at fixed $L$ is one-to-one; for double-spin
binaries, a range of $J$ are allowed at each fixed $\xi$.  To guide
the eye in the plots that follow, we evaluate and show this
``allowed'' region in the $J\xi$-plane for the chosen source
parameters $m_1,m_2,\chi_1,\chi_2$ in Figure~\ref{fig:BoundaryJXi}.   
We will use $J_{\rm r0}(\xi),J_{\rm r\pi}(\xi)$ to denote the maximum
and minimum values of $J$ for a given $\xi$.  Points on these two
curves, including the injections described above, are all resonant
binaries: binaries with $J=J_{\rm r\pi}(\xi)$ belong to the
$\Delta\Phi=\pi$ resonance, while those with $J=J_{\rm r0}(\xi)$
belong to the $\Delta\Phi=0$ resonance.
Additionally, as described in Refs.~\cite{2015PhRvL.114h1103K} and \cite{2015PhRvD..92f4016G}, these parameters facilitate \emph{morphological
  classification}, subdividing the parameter space ${\cal H}$ into three disjoint regions ${\cal H}_0,{\cal
  H}_C,{\cal H}_\pi$, set by the (nondissipative)
dynamics at $f_{\rm ref}=100\unit{Hz}$.  Specifically, ${\cal H}_0$ and
${\cal H}_\pi$ are the systems whose spins are librating about the $\Delta\Phi=0$ and $\Delta\Phi=\pi$ resonance, respectively; ${\cal
  H}_C$ are the remaining, circulating binaries.    
Geometrically,  these three regions are separated by configurations where either $\mathbf{S}_1$ or $\mathbf{S}_2$ are parallel to
$\mathbf{L}$ at some point during one precession cycle. 
From a computational point of view, motivated by
\cite{2015PhRvL.114h1103K,2015PhRvD..92f4016G}, at each $m_1,m_2,\chi_1,\chi_2$ we
determine the morphology from the values of $J$ and $\xi$ at
$100~\unit{Hz}$.
We define
$J_{c0}(\xi),J_{c\pi}(\xi)$ as the two values of $J$ which allow either $\mathbf{S}_{1}$ or $\mathbf{S}_{2}$ to be parallel to $\mathbf{L}$ at some point
during their precession cycle, ordered as  $J_{\rm r\pi}(\xi)<J_{c\pi}(\xi)<J_{c0}(\xi)<J_{\rm r0}(\xi)$.
Because these surfaces separate the different morphologies, we identify the appropriate morphology simply by comparing
 $J$ to $J_{\rm c0,c\pi}(\xi)$: binaries with  $J<J_{\rm c\pi}(\xi)$
 belong to
${\cal H}_\pi$; binaries with  $J>J_{\rm c0}(\xi)$ 
belong to ${\cal H}_0$; and binaries with $J_{\rm c\pi} <J<J_{\rm c 0}$ belong to ${\cal H}_C$.  
These boundaries are illustrated in Figure \ref{fig:BoundaryJXi}.

We apply morphological
decomposition to the list of posterior samples, classifying the fraction $p(A)$ of samples that
are consistent with each morphology $A={\cal H}_0$, ${\cal H}_\pi$, and ${\cal H}_C$, respectively.  
As described in previous work (see in
particular~\cite{2015PhRvL.114h1103K,2015PhRvD..92f4016G}), while
librating binaries occur frequently near merger for comparable-mass
binaries, a relatively small amount of the possible parameter space
describing all compact binaries corresponds to librating orbits
trapped near a resonance (${\cal H}_{\pi,0}$), as opposed to
circulating orbits (${\cal
  H}_C$).

Using the prior parameter distribution described below
Eq.~(\ref{eq:Bayes}),\footnote{To ensure consistency, we draw samples
  from the prior by running the parameter estimation code with input
  $h(t)=0$.  While we are forced to adopt the ad-hoc choices of
  Eq.~(\ref{eq:prior}) for prior probabilities to be consistent with
  the mass and mass ratio distribution described previously and used
  by convention in \textsc{lalinference}, we emphasize that those
  probability distributions were chosen arbitrarily to avoid the
  impression of bias in mass.  Nature likely favors other mass
  distributions
  \cite{2012ApJ...759...52D,2013ApJ...779...72D,2015ApJ...806..263D},
  which may further enhance the prior probability that non-circulating
  morphologies occur. }
we find prior
 probabilities $p_{\rm prior}({\cal H}_{0,\pi,C})$
\begin{align}
\begin{aligned}
 &p_{\rm prior}({\cal H}_0) =  0.118\,; \\  
 &p_{\rm prior}({\cal H}_\pi) = 0.049\,; \\
\label{eq:prior}
 &p_{\rm prior}({\cal H}_C) = 0.832\,.
 \end{aligned}
 \end{align}
The prior also implies a marginal distribution $p(J,\xi)$, shown in the right panel of Figure \ref{fig:BoundaryJXi}.  The
marginal distribution has several salient features, most notably a peak near $(J,\xi)\simeq  (0.6, 0)$ arising for
example from randomly-oriented spins in unequal-mass binary black holes.  Extending away from that peak are three
wedges, corresponding to binaries (a) with spins increasingly aligned with $\mathbf{L}$, extending upward toward
$J\simeq 1$ and $\xi \simeq 1$; (b) with spins increasingly anti-aligned with $\mathbf{L}$, extending downward towards 
$\xi \simeq -1$; and (c) a region extending to the left along $\xi \simeq 0$ with small  $J$, corresponding for example
to small mass ratios $q$ and small primary spins $\chi_1$.

The posterior distribution can also be parametrized using the two
frequencies implied by the spin precession equations,
evaluated (to avoid ambiguity) using a reference orbital frequency:
here $f_{\rm ref}=2f_{\rm orb}=100\unit{Hz}$.  These two timescales
appear directly in the spin and orbital dynamics, and hence are
reflected in the waveform.
Rather than adopt a precession-phase dependent definition of these two
timescales via, e.g., the vector $\mathbf{\Omega}_{i}$ that appear in
the spin precession equations
$d\mathbf{S}_i/dt=\mathbf{\Omega}_i\times \mathbf{S}_i$, following
\citet{2015PhRvL.114h1103K,2015PhRvD..92f4016G} we use an analytic,
explicitly doubly periodic solution for the nondissipative precession
dynamics to identify a precession timescale $\tau$ for the relative
spin motion [defined below Eq.~(8) in \cite{2015PhRvL.114h1103K}]
and a precession frequency
$\left<\Omega_z\right>$ for the motion of $\mathbf{L}$ around
$\mathbf{J}$.  Relative to their notation,
$\left<\Omega_z\right>\equiv\alpha/\tau$, where $\alpha$ is the
precession angle of $\mathbf{L}$ around $\mathbf{J}$ in one spin
precession period $\tau$; see Appendix \ref{ap:alpha} for details.

\section{\label{sec:Results}Results}

\subsection{Robust morphological classification}

\begin{figure*}\centering
 \vspace{-1cm}
\includegraphics[width=\textwidth]{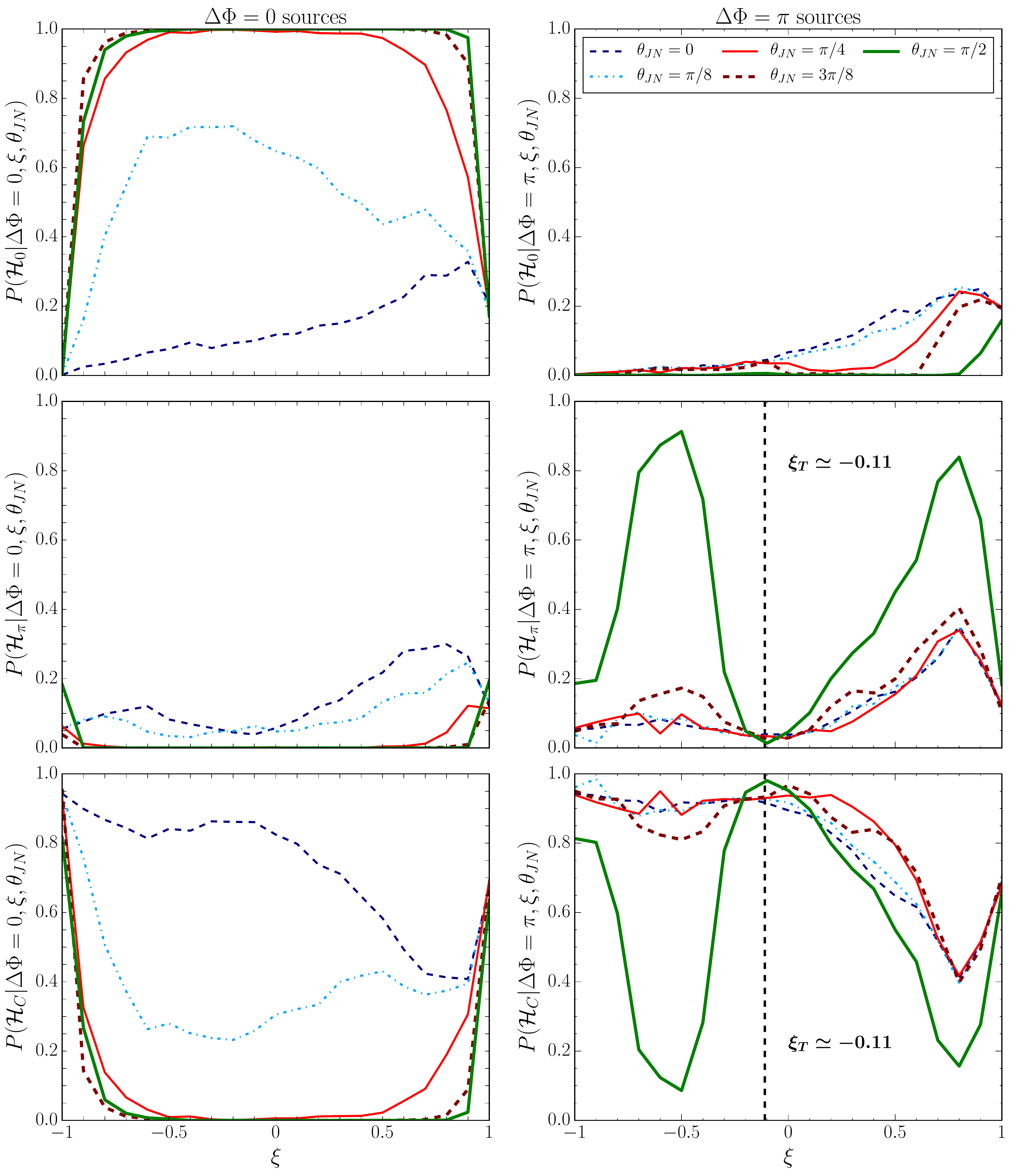}

\caption{
\label{fig:ProbabilityVersusParameters}\textbf{Constraining morphology versus inclination: the posterior.} Classification probability versus $\xi$ for different values of $\theta_{JN}$ at fixed ${\rm SNR}=20$. %
Panels in this plot are organized as a matrix: sources ($\Delta\Phi=0$ or $\Delta\Phi=\pi$) are columns; the
fractions of the posterior in each morphology (${\cal H}_{0},{\cal H}_\pi,{\cal H}_C$) are rows; and each panel shows the
corresponding probability evaluated for sources at different values of $\xi$.  For example, the bottom-left panel shows
the probability $p({\cal H}_C|\Delta\Phi=0)$ versus $\xi$ and (in different line styles) versus $\theta_{JN}$. 
Nearly aligned binaries ($\xi \simeq 1,-1$ or $\Delta\Phi=\pi$ and $\xi=\xic\simeq \xicVal$) and nearly face-on
binaries ($\theta_{JN}\simeq 0$) cannot be reliably classified (top left and middle right).  A source with
$\Delta\Phi=0$ can be reliably classified in the ${\cal H}_0$ morphology  (top left) unless face-on (for our simulations, $\xi \approx
\pm 1$);
however, a source
with $\Delta\Phi=\pi$ can be reliably classified in the ${\cal H}_\pi$ morphology only if $\theta_{JN}\simeq \pi/2$
 and $\xi \ne -1,1,\xic$ (green curve in middle right).  
In other words, cases where the probabilities vary significantly with $\xi$ reflect  how the misalignment angles between the various angular momenta change
versus $\xi$: compare to Figure \ref{fig:Recap}.  
Due to the relatively small number ($N\simeq 2000$) of posterior samples used, these
probabilities have a few percent (binomially distributed)  sampling uncertainty. 
}
\end{figure*}

\begin{figure}
\hspace{-.5cm}\includegraphics[width=\columnwidth]{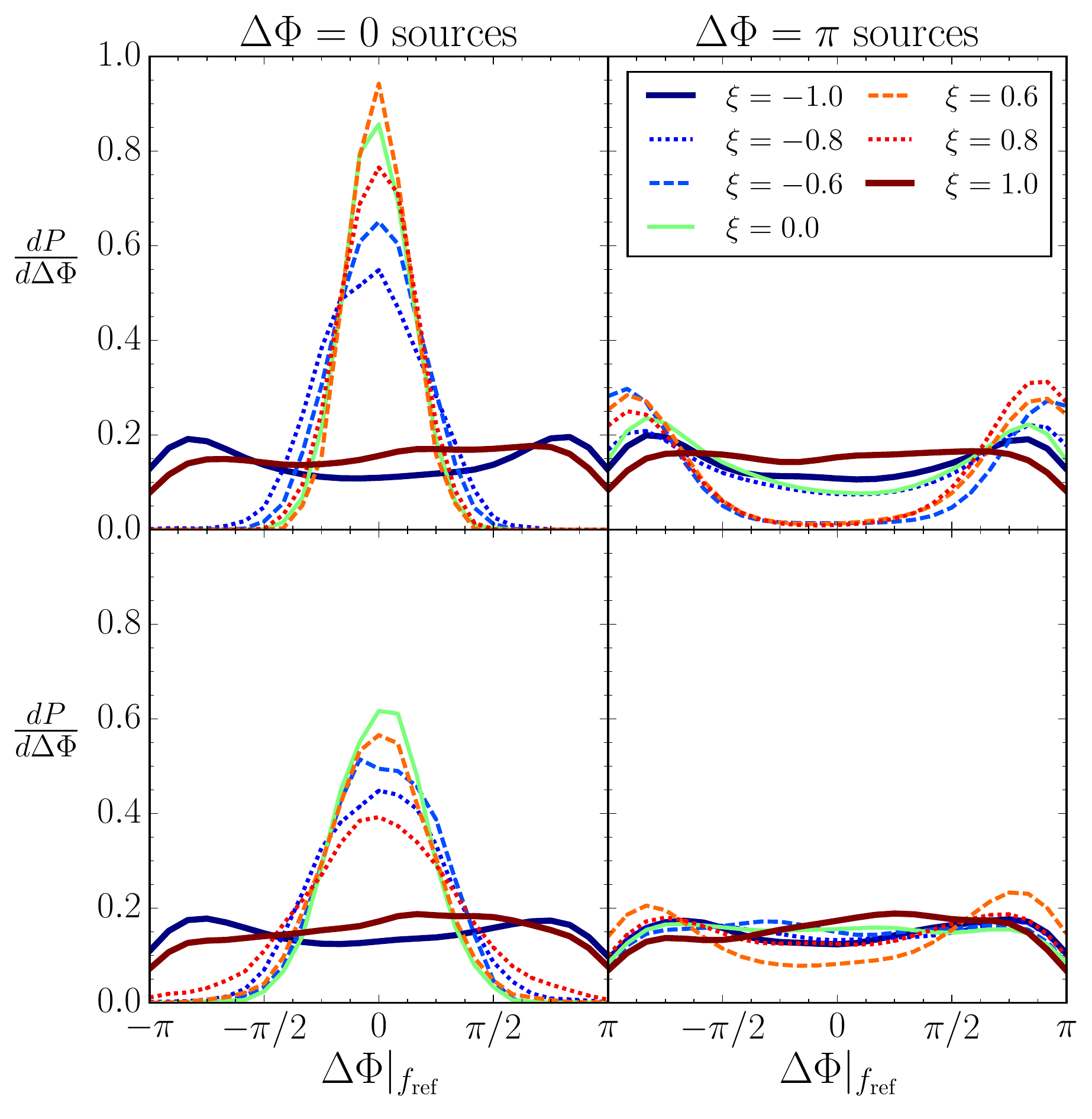}
\caption{\label{fig:ConstrainDeltaPhi}\textbf{Constraining
    $\Delta\Phi$ at  $100~\unit{Hz}$.} %
  Probability distribution $dP/d\Delta\Phi$  of the phase angle
  $\Delta \Phi$ between the spins can be reliably measured for
  $\Delta \Phi\simeq 0$.  Plot shows the posterior distribution for
  $\Delta\Phi$ between the spins using spin vectors at the reference
  frequency $f_{\rm ref}=100~{\rm Hz}$, given data containing sources with
  $\Delta\Phi=0$ (left column) or $\Delta\Phi=\pi$ (right column),
  with either $\theta_{JN}=\pi/2$ (top) or $\theta_{JN}=\pi/4$
  (bottom).  In all panels, colors indicate different values of $\xi$,
  as shown in the legend.  This figure shows that the posterior
  distribution in the intrinsic parameter $\Delta\Phi$ depends
  strongly on the extrinsic parameter $\theta_{JN}$ and weakly on the
  intrinsic parameter $\xi$, except for nearly aligned systems
  $\xi \simeq \pm 1$. }
\end{figure}

\begin{figure*}\centering
\includegraphics[width=\textwidth]{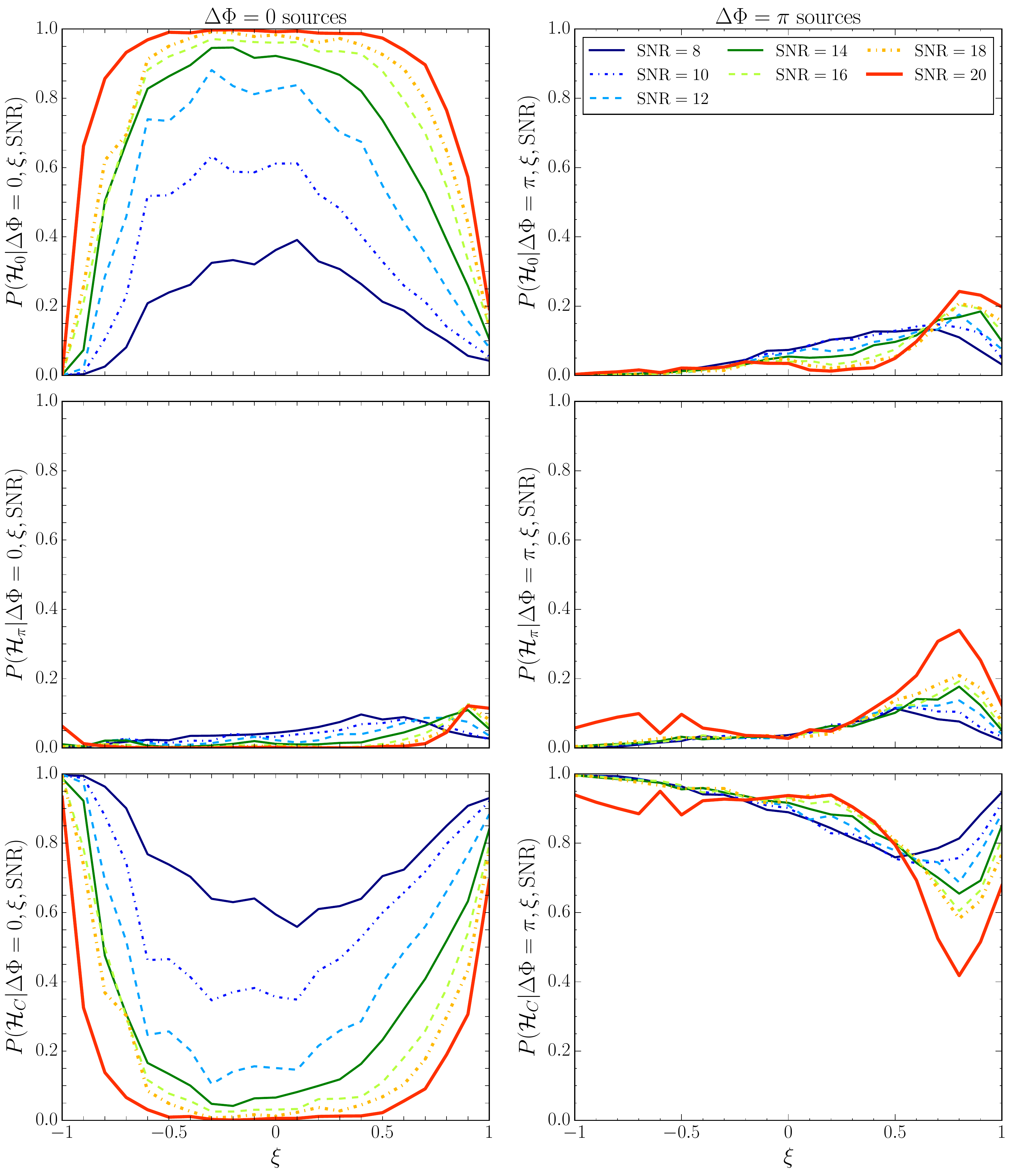}

\caption{\label{fig:SNRVersusParameters}\textbf{Constraining morphology versus SNR.}
Plot of the classification probability versus $\xi$ for different choices of SNR and $\theta_{JN}=\pi/4$.
The panels in this figure are organized as in Figure \ref{fig:ProbabilityVersusParameters}.
  As described in Appendix \ref{ap:details}, the ${\rm SNR}=20$ results (red curve) were produced using a lower starting frequency than
  results for SNR$<20$; this discrepancy is responsible for the slight difference in trend between the right  panels'
  results for SNR=20 versus results for SNR$<20$.
}
\end{figure*}

Let us first consider sources with $\Delta\Phi=0$. Except for binaries
that are either face-on ($\theta_{JN}=0$)
or have both spins nearly aligned with $\mathbf{L}$ ($|\cos\theta_{LS_1}|\simeq 1$ and
$|\cos\theta_{LS_2}|\simeq 1$), our calculations show that the
posterior probability $p({\cal H}_0)$ is a good indicator of which
resonant family the source came from, for almost all source
orientations and $\xi$, and for astrophysically plausible source
amplitudes.  For example, the left panel of Figure
\ref{fig:ProbabilityVersusParameters}
shows the posterior probability
$p({\cal H}_0|\Delta\Phi=0,\xi,\theta_{JN})$: the fraction of
posterior probability implying the source librates about the
$\Delta\Phi=0$ resonance, as a function of the true source parameters
$(\xi,\theta_{JN})$ at a fixed source amplitude ${\rm SNR}=20$.
This probability is large (and nearly unity) for almost all
$(\xi, \theta_{JN})$.  For binaries at or near the $\Delta\Phi=0$
resonance, we can confidently classify them in their corresponding
resonant regime.  Equally striking, the center-left panel shows that
the probability of misidentification in the opposite resonance is
nearly zero: $p({\cal H}_\pi|\Delta\Phi=0,\xi,\theta_{JN})\simeq 0$.
These high probabilities should be sharply contrasted with the small
prior probability associated with the $\Delta\Phi=0$ resonance
[$p_{\rm prior}({\cal H}_0)=0.118$], as the majority of phase space is
associated with circulating binaries.

The right panels of Figure \ref{fig:ProbabilityVersusParameters} show
our corresponding but qualitatively different results using source
binaries from the $\Delta\Phi=\pi$ resonant family.
Except for specific configurations, our calculations show that the
posterior probability favors the circulating morphology and disfavors the ``incorrect''
classification. This follows from geometrical considerations: $\Delta\Phi=\pi$ sources
are more aligned and precess less, making them more difficult to distinguish from non-precessing
sources.
The probability of ``correct'' classification
$p({\cal H}_\pi|\Delta\Phi=\pi,\xi,\theta_{JN})$ shown in the
middle-right panel decreases significantly at
$\xi=\xic \simeq \xicVal$. This value of $\xi$ corresponds to a special configuration
featured by $\Delta\Phi=\pi$ sources, with $\cos\theta_{LS_1} = -1$ and
$\cos \theta_{LS_2}=1$; see, e.g., our Figure \ref{fig:Recap} or
Figure 2 of \cite{2014PhRvD..89l4025G}.  For our injection
masses and spin magnitudes, this ``aligned''
configuration occurs at $\xi=\xic \simeq \xicVal$, as explained
earlier.\footnote{The spin-aligned configuration at $\xi=\xi_T$ on the $\Delta\Phi=\pi$ family arises because the down/up configuration is a stable
  fixed point of the spin precession equations; by contrast, for our masses, spin magnitudes, and frequencies, the up/down configuration is unstable and not a member of
  the $\Delta\Phi=0$ family   \cite{2015PhRvL.115n1102G}.  When the up/down configuration is stable, the point with $\cos \theta_1 = +1$ and $\cos \theta_2 = -1$ will belong to the $\Delta\Phi=0$ resonant family and we anticipate similarly diminished precession amplitudes in its vicinity leading to a similarly difficult challenge in identifying the morphology.}
Resonant binaries at and near $\xic$ have spins that barely precess,
being almost parallel to 
$\mathbf{L}$.  In other words,
just like the two limits $\xi\rightarrow 1$ and $\xi\rightarrow -1$,
when both spins are coaligned with each other and $\mathbf{L}$,
resonant sources with $\Delta\Phi=\pi$ and $\xi\simeq \xic$ have
almost indistinguishable dynamics from a \emph{non-precessing} binary.
As seen in Figure \ref{fig:ProbabilityVersusParameters}, for sources
with $\xi\simeq \xic$ and $\Delta\Phi=\pi$ the morphology is only
weakly constrained by observation:
$p({\cal H}_\pi|\Delta\Phi=\pi,\xicVal,\theta_{JN})$ is small,
comparable to the prior (see Sec.~\ref{sec:KLdivergence} and
Fig.~\ref{fig:KLdivergence} below).
Second, the probability for correct classification of a
$\Delta\Phi=\pi$ source is often significantly smaller than the
corresponding $\Delta\Phi=0$ result.
As emphasized in \cite{2014PhRvD..89l4025G}, the $\Delta\Phi=\pi$ and
$\Delta\Phi=0$ resonant families have qualitatively different dynamics
and GW signals: $\Delta\Phi=\pi$ resonances undergo substantially less
precession of $\mathbf{L}$, and hence have less modulated GWs.  As a
result, at fixed signal amplitude, GW measurements of sources from the
$\Delta\Phi=\pi$ resonance should be less effective at measuring
double-spin physics like the morphology, implying the lower
probabilities
$p({\cal H}_\pi|\Delta\Phi=\pi,\xi,\theta_{JN}) \lesssim p({\cal
  H}_0|\Delta\Phi=0,\xi,\theta_{JN}) $
seen in the corresponding panels of Figure
\ref{fig:ProbabilityVersusParameters}.
To qualitatively corroborate this analysis, we also examined the posterior for $\Delta\Phi$; an example appears in
Figure~\ref{fig:ConstrainDeltaPhi}.
In most cases, $\Delta\Phi$ could be weakly  constrained to be $\simeq 0$ or $\simeq \pi$.  
By definition, the region
${\cal H}_0$ corresponds to libration about the $\Delta\Phi=0$ resonance.  Because the phase-space trajectories in this region explicitly cannot extend to
$\Delta\Phi=\pi$ \cite{2015PhRvL.114h1103K,2015PhRvD..92f4016G}, posterior samples in this region explicitly disfavor $\Delta\Phi \simeq \pi$; in fact,  depending
on $\xi$, the allowed region of $\Delta\Phi$ can be narrow.%
\footnote{At sufficiently high signal amplitudes,  phase
  angles like $\Delta\Phi$ can be measured directly, independent of
  how rapidly they evolve with frequency.  For example, the phase
  angle $\phi_{JL}$ can be estimated to within an angle of order
  $2\pi/{\rm SNR}$, with a coefficient that depends on the line of
  sight ($\theta_{JN}$) and the amplitude of precession of $\mathbf{L}$ 
  around $\mathbf{J}$ (e.g., the characteristic angle between
  $\mathbf{L}$ and $\mathbf{J}$): see, e.g.,
  \cite{gwastro-mergers-HeeSuk-CompareToPE-Precessing}.
}  
To summarize, our ability to confidently classify source morphology to ${\cal H}_{0,\pi}$ implies $\Delta\Phi$ can be
constrained away from $\pi,0$, respectively.   
That said, morphological classification is more robust and qualitatively different than measurements of $\Delta\Phi$;
for example, while $\Delta\Phi$ changes on the
precession timescale, a binary's 
morphology only changes on the inspiral timescale. 
All of our results depend on the signal amplitude, here fixed at
${\rm SNR}=20$.  Repeating parts of our study with much stronger
signals (${\rm SNR}\simeq 50$), we have found dramatically improved
morphological classification, even for sources with $\Delta\Phi=\pi$.  
Conversely,  our ability to reliably classify morphologies gradually
degrades at lower source amplitude, as shown in Figure~\ref{fig:SNRVersusParameters}.  

\begin{figure*}
\begin{tabular}{c}
	\hspace{-2.7cm}\includegraphics[width=0.905\textwidth]{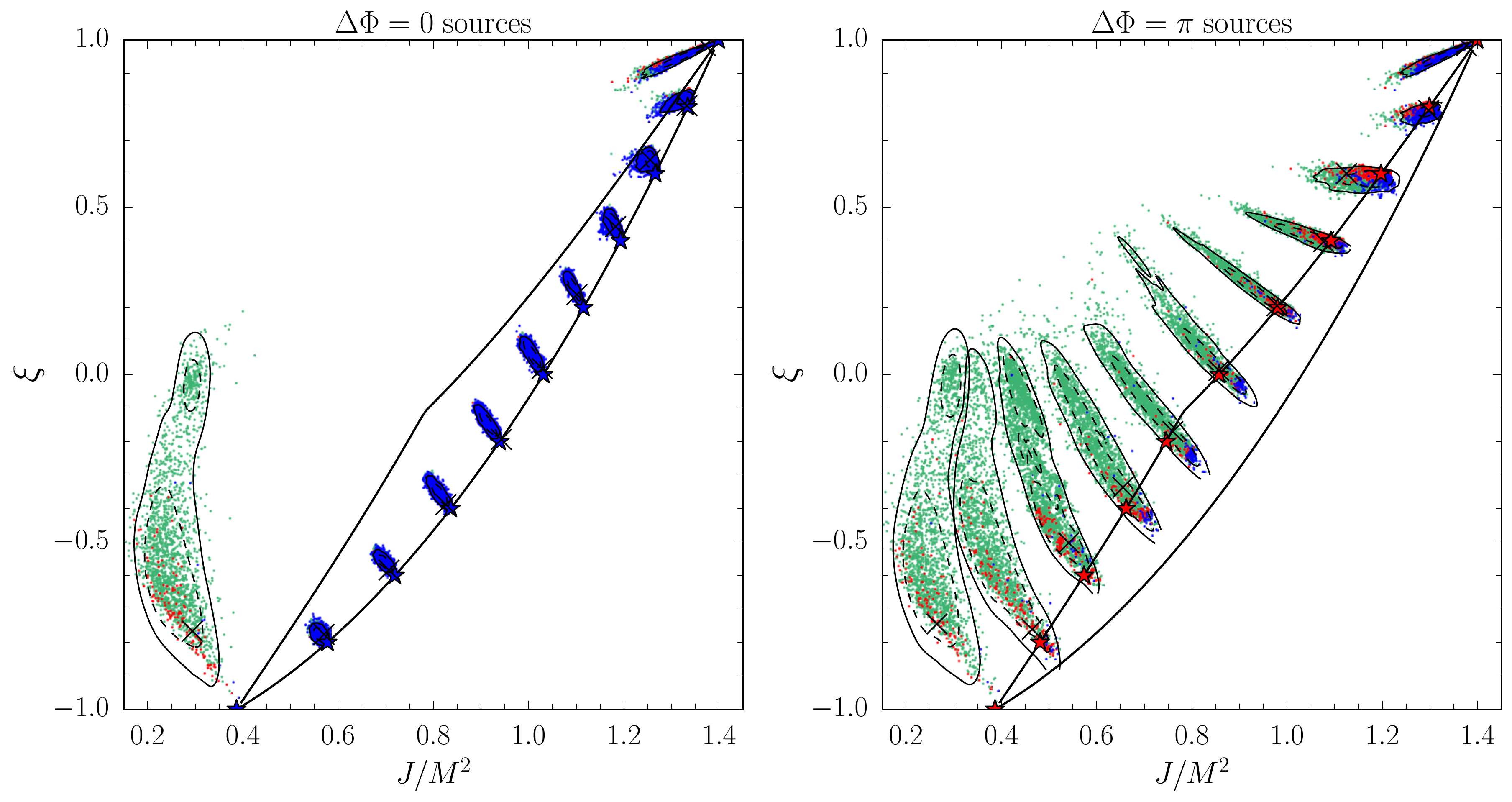}\\
	\hspace{-1cm}\includegraphics[width=\textwidth]{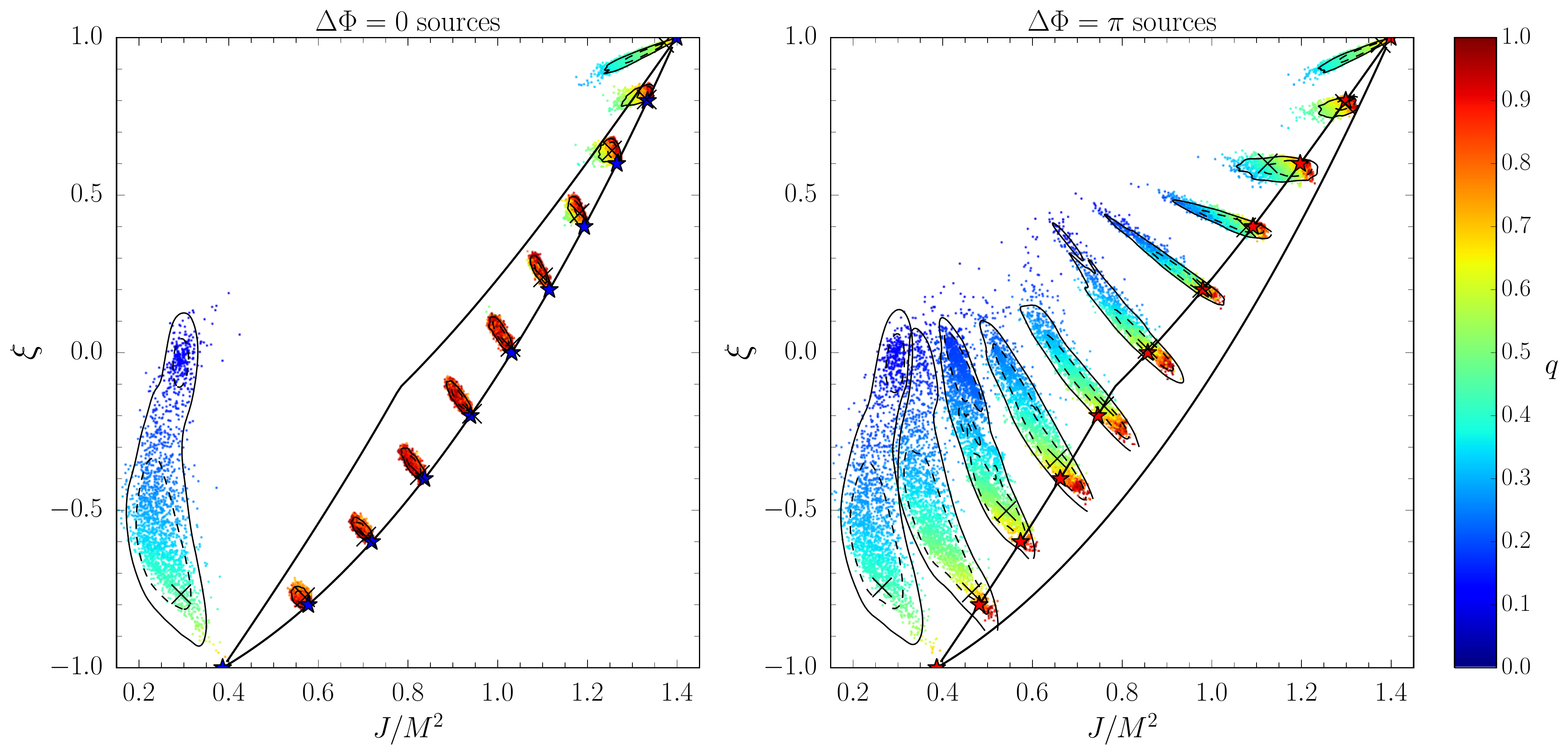}
\end{tabular}
\caption{\label{fig:ConstrainJXi}\textbf{Constraining $(J,\,\xi)$.}
  The parameters $J$ and $\xi$
can be independently constrained to a narrow volume 
of parameter space.  Posterior distributions from distinct simulations in the $J\xi$-plane for
$\theta_{JN}=\pi/4$ and ${\rm SNR}=20$, for source binaries with $\Delta\Phi=0$ (left) and $\Delta\Phi=\pi$ (right).
The actual value from each simulation is marked with a star;
 the maximum-likelihood estimate is marked by a black cross.
To guide the eye, the allowed region for $J$ and $\xi$ for the true source parameters is superimposed (thick black line) %
as in Figure \ref{fig:BoundaryJXi}. As the allowed region depends on masses and spins, not all points in the posterior
lie in this region;  see~\cite{2015PhRvD..92f4016G} for details. Solid and dashed black lines show the  95\%  and 67\% confidence
intervals for each set of posterior samples from distinct simulations. \emph{Top panels:} Samples
are colored according to morphology (blue: librating around $\Delta\Phi=0$, red: librating around $\Delta\Phi=\pi$,
green: circulating). 
\emph{Bottom panels:} 
Samples are colored according to the posterior mass
ratio $q$.  
 In our simulations, binaries without sufficiently strong precession-induced
modulation to enable  tight constraints on $q$ are equally poorly constrained in morphology:  our ability to classify
morphology correlates with our ability to constrain $q$.
}
\end{figure*}

\subsection{Using conserved quantities as  coordinates to describe double-spin measurements}
On physical grounds, we expect GW measurements to best constrain
quantities that are constant on the longest timescales.  In a two-spin
system, while the specific component spins and orbital angular momenta
depend effectively on a precession phase choice, recent analytic
work~\cite{2015PhRvL.114h1103K,2015PhRvD..92f4016G} has identified
natural quantities, conserved on the precession timescale, to
characterize the orbit: $(J,\,\xi)$.

Figure \ref{fig:ConstrainDeltaPhi} shows posterior probability distributions for $\Delta\Phi$, for several fiducial source events; Figure
\ref{fig:ConstrainJXi} shows corresponding posterior distributions for $(J,\xi)$.   
To guide the eye, in Figure \ref{fig:ConstrainJXi}, a thick black line shows the allowed region for $J,\xi$ assuming all
mass and spin parameters are equal to the source parameters, as in Figure \ref{fig:BoundaryJXi}.
First and foremost, as noted above, the measurements shown in Figure \ref{fig:ConstrainDeltaPhi} immediately reveal that $\Delta\Phi$ at $f=100\unit{Hz}$ can be
measured for many  binaries which do not have $\mathbf{S}_{1,2}$ parallel to $\mathbf{L}$.
Though nominally an
intrinsic parameter tied to the phase of the spins' relative precession cycle,  the parameter  $\Delta\Phi$ is limited by the range of precession dynamics  allowed by the system at $f_{\rm ref}$. 
Second, the posterior in $(J,\xi)$ is highly nongaussian, as shown in
Figure \ref{fig:ConstrainJXi}.
Finally,  precession-induced modulation is known to break all degeneracies and enable strong constraints on many
parameters \cite{2006PhRvD..74l2001L,2008ApJ...688L..61V,2011PhRvD..84b2002L,gwastro-mergers-HeeSuk-FisherMatrixWithAmplitudeCorrections,gwastro-mergers-HeeSuk-CompareToPE-Precessing,2015PhRvD..92d4056M,LIGO-CBC-S6-PE}.   As shown by the
bottom panel of Figure \ref{fig:ConstrainJXi}, our ability to constrain the posterior distributions in $(J,\xi)$ and in
mass ratio is correlated.

\begin{figure*}[!htpb]
	\includegraphics[width=.87\textwidth]{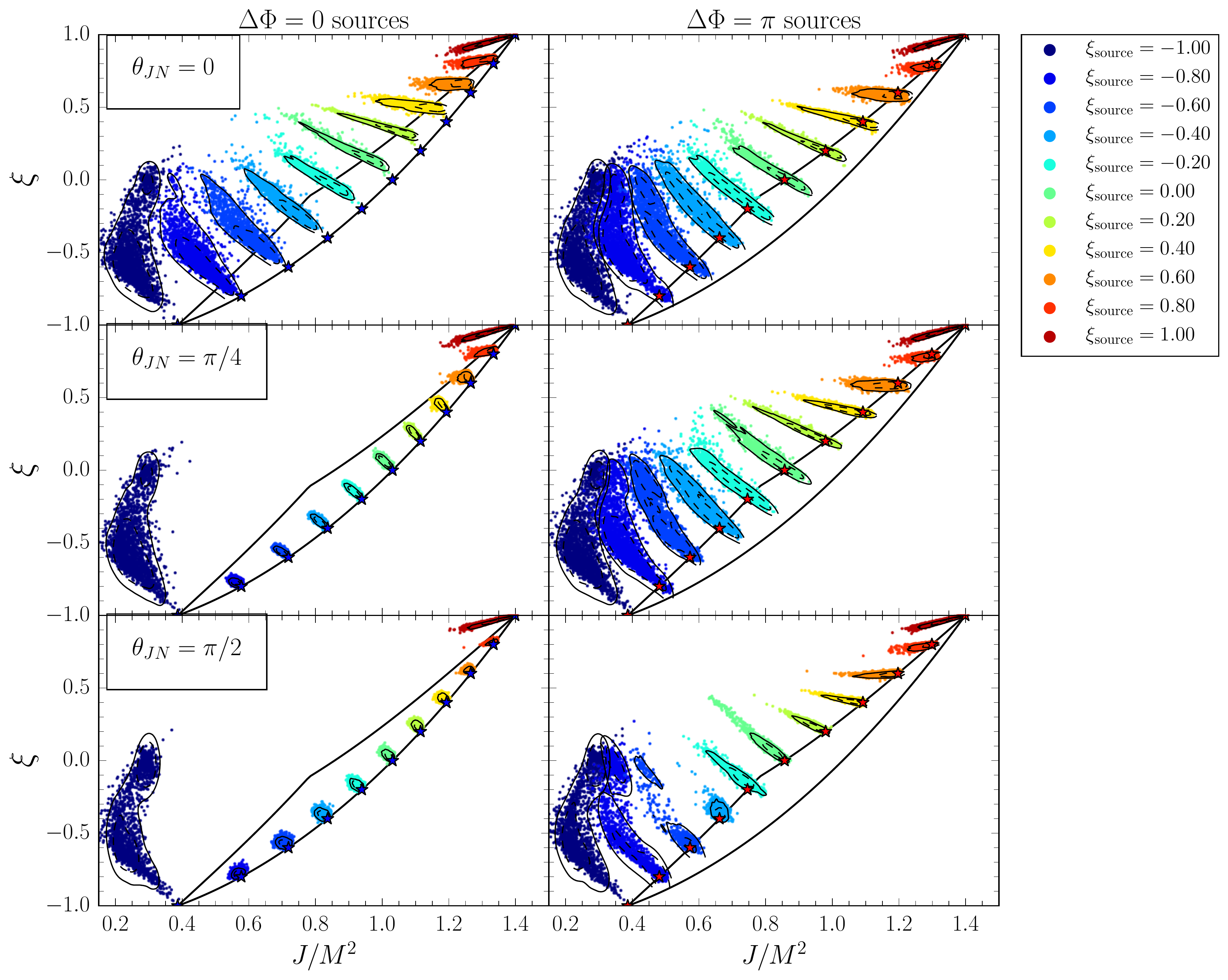}
        \caption{\label{fig:multipanelThetaJN}\textbf{
        Constraining $(J,\xi)$: dependence on $\theta_{JN}$.}  Posteriors from
          different simulations in the $J,\xi$ plane for fixed ${\rm SNR}$ and different
          $\theta_{JN}$, increasing from top ($\theta_{JN}=0$) to
          bottom ($\theta_{JN}=\pi/2$). Points are colored according to the $\xi$ values
          of the simulations; in this figure, all sources have SNR=20.
}
\end{figure*}

Figure \ref{fig:multipanelThetaJN} shows the posterior distributions for sources with different orientation
$\theta_{JN}$ relative to the line of sight.   Similar to the change in morphology constraints with $\theta_{JN}$, the extent of the
posterior distribution in the $J\xi$-plane increases (roughly) as  $\sin \theta_{JN}$.  Nearly face-on sources have little 
precession-induced modulation in their GW signal, and therefore measurements  cannot as reliably infer $J,\xi$.

\begin{figure*}[!htpb]
	\includegraphics[width=.87\textwidth]{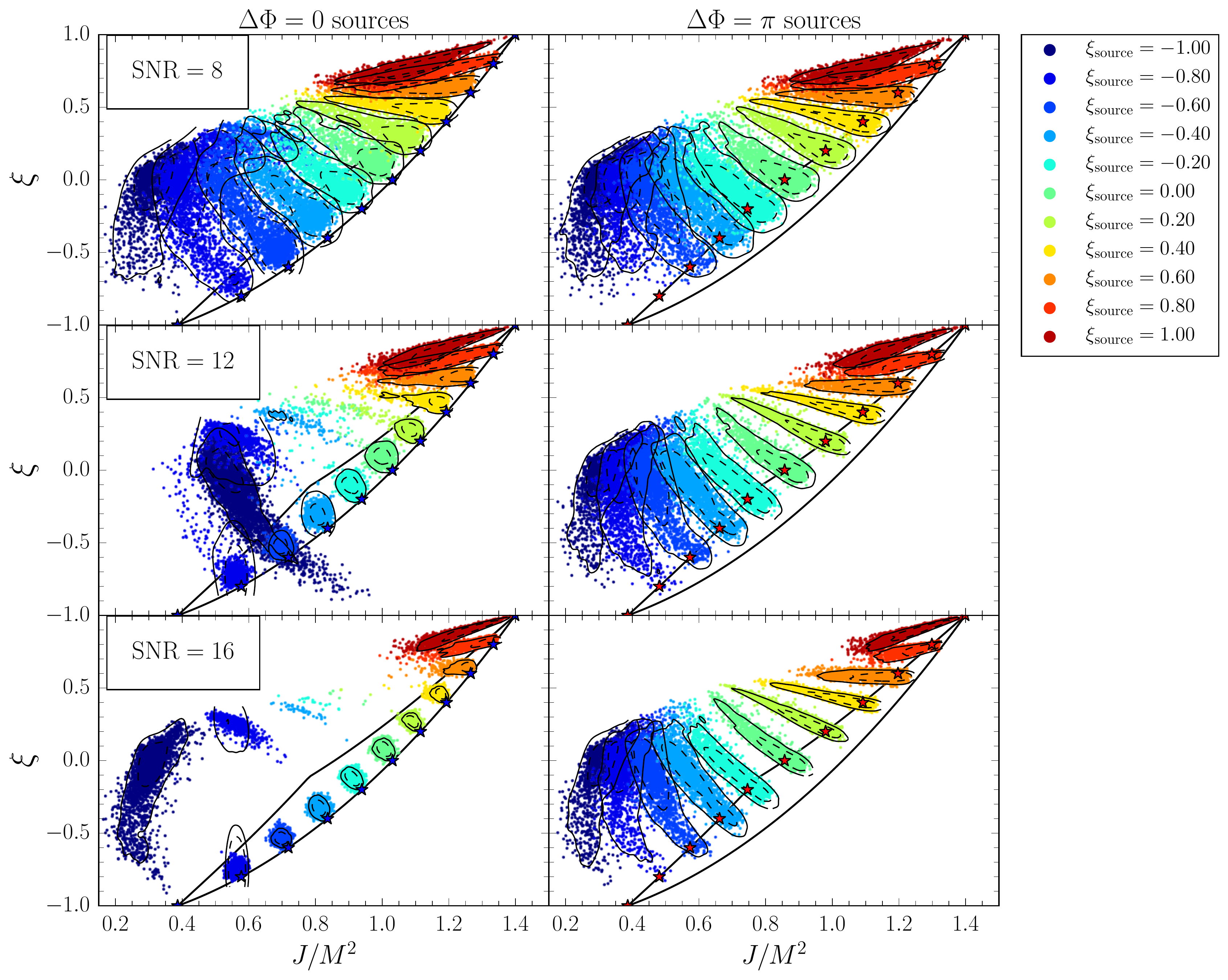}

        \caption{\label{fig:multipanelSNR}\textbf{
        Constraining
            $(J,\xi)$: dependence on SNR.}  Posteriors from different
          simulations in the $J\xi$-plane for fixed $\theta_{JN}=\pi/4$ and different ${\rm SNR}$,
          increasing from top (${\rm SNR}=8$) to bottom
          (${\rm SNR}=16$). Points are
          colored according to the simulation's $\xi$ value.
          }
\end{figure*}

Finally, Figure \ref{fig:multipanelSNR} shows the posterior distributions for sources with different network amplitudes (SNR).
Similar to the change in morphology constraints with SNR, the extent of the
posterior distribution in the $J\xi$-plane increases as the source SNR decreases.

Comparing our results to the prior $J,\xi$ distribution shown in Figure \ref{fig:BoundaryJXi} suggests the wide prior
convention has a significant impact on the posterior distribution, ``pulling'' the posterior away from the region
most strongly supported by the data.   We anticipate that astrophysically motivated priors which favor larger spins and more comparable mass
ratios should produce tighter constraints on $J,\xi$ and morphology.

\subsection{Natural timescales as coordinates for double-spin measurements}

Rather than using the conserved constants $(J,\xi)$ to characterize
spin precession, we can use the two natural precession frequencies
$(\left<\Omega_z/(2\pi)\right>,1/\tau)$ introduced
in~\cite{2015PhRvL.114h1103K,2015PhRvD..92f4016G} and reviewed in
Appendix~\ref{ap:alpha}.  
Unlike the instantaneous precession vectors which appear in
$d\mathbf{S}_{1,2}/dt$, these expressions are independent of
precession phase.
Moreover, even for high-symmetry configurations like nearly aligned
spins and post-Newtonian resonances, these timescales -- corresponding
to the \emph{rate} of precessional modulation -- are except for one special case well-defined and
finite,\footnote{As described in \citet{2015PhRvL.115n1102G}, for certain seperations a binary with both spins parallel to
  $\mathbf{L}$, with the more massive ``up'' and the less massive ``down'', is unstable to spin precession.  When the
  instability occurs, a single point in the interior of the $J,\xi$ region has $\tau=\infty$. } though the associated \emph{amplitudes} can be zero.  For
example, the frequency $\left<\Omega_z\right>\equiv\alpha/\tau$ is
nonzero for binaries with $\mathbf{L,S_1,S_2}$ all parallel,
as can most easily be demonstrated in the special case of a single
precessing spin \cite{1994PhRvD..49.6274A}.  Similarly, even though no modulation on the timescale takes place in post-Newtonian
resonances, where spins and $\mathbf{L}$ remain coplanar, the spin precession timescale $\tau$ is well-defined and
nonzero \cite{2015PhRvD..92f4016G}.

As described in Appendix \ref{ap:alpha}, due to coordinate ambiguity
when $\mathbf{J}$ and $\mathbf{L}$ are parallel, the angle $\alpha$
can change by $2\pi$ on two distinct surfaces in the $J\xi$-plane, but
is otherwise continuous.
The function $\tau$ is continuous.  Because the coordinate
transformation from $(J,\xi)$ to
$(\left<\Omega_z\right>/(2\pi),1/\tau)$ has two simple
discontinuities, the connected posterior distributions in $(J,\xi)$
seen above can map to two regions in
$(\left<\Omega_z\right>/(2\pi),1/\tau)$ when the posterior distribution
is sufficiently poorly constrained or close to these critical
surfaces.
More broadly, the transformation between  $J,\xi$ and $\Omega_z,\tau$ may not always be globally one-to-one: the
transformation might lose some information.  That said, because these timescales occur in the dynamics and hence
waveform, their imprint should appear in the posterior distribution, as with single-spin binaries \cite{gwastro-mergers-HeeSuk-CompareToPE-Precessing}.

\begin{figure*}[p]
\begin{tabular}{c}
\includegraphics[width=\textwidth]{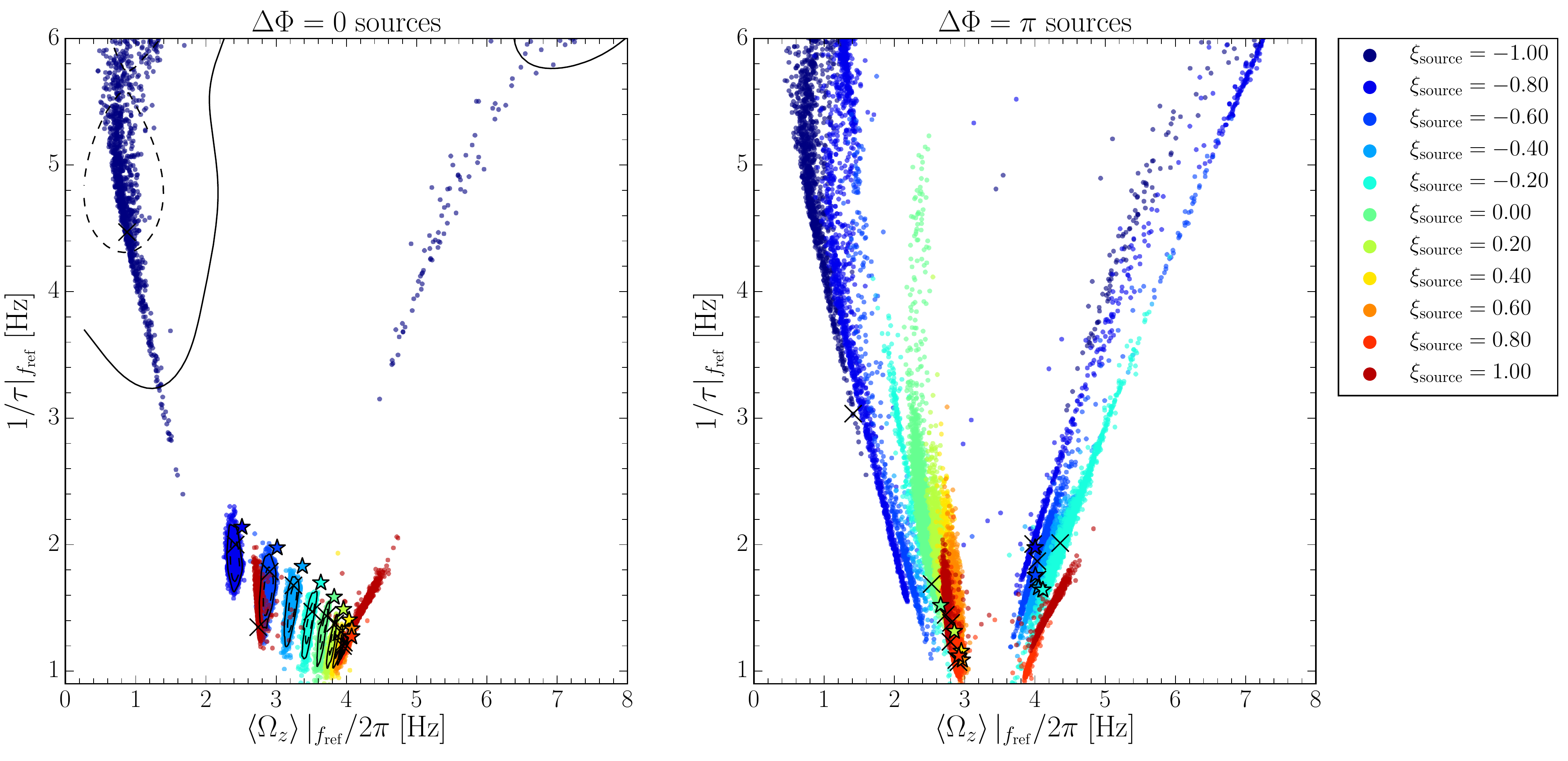}\\
\hspace{-2.8cm}\includegraphics[width=0.87\textwidth]{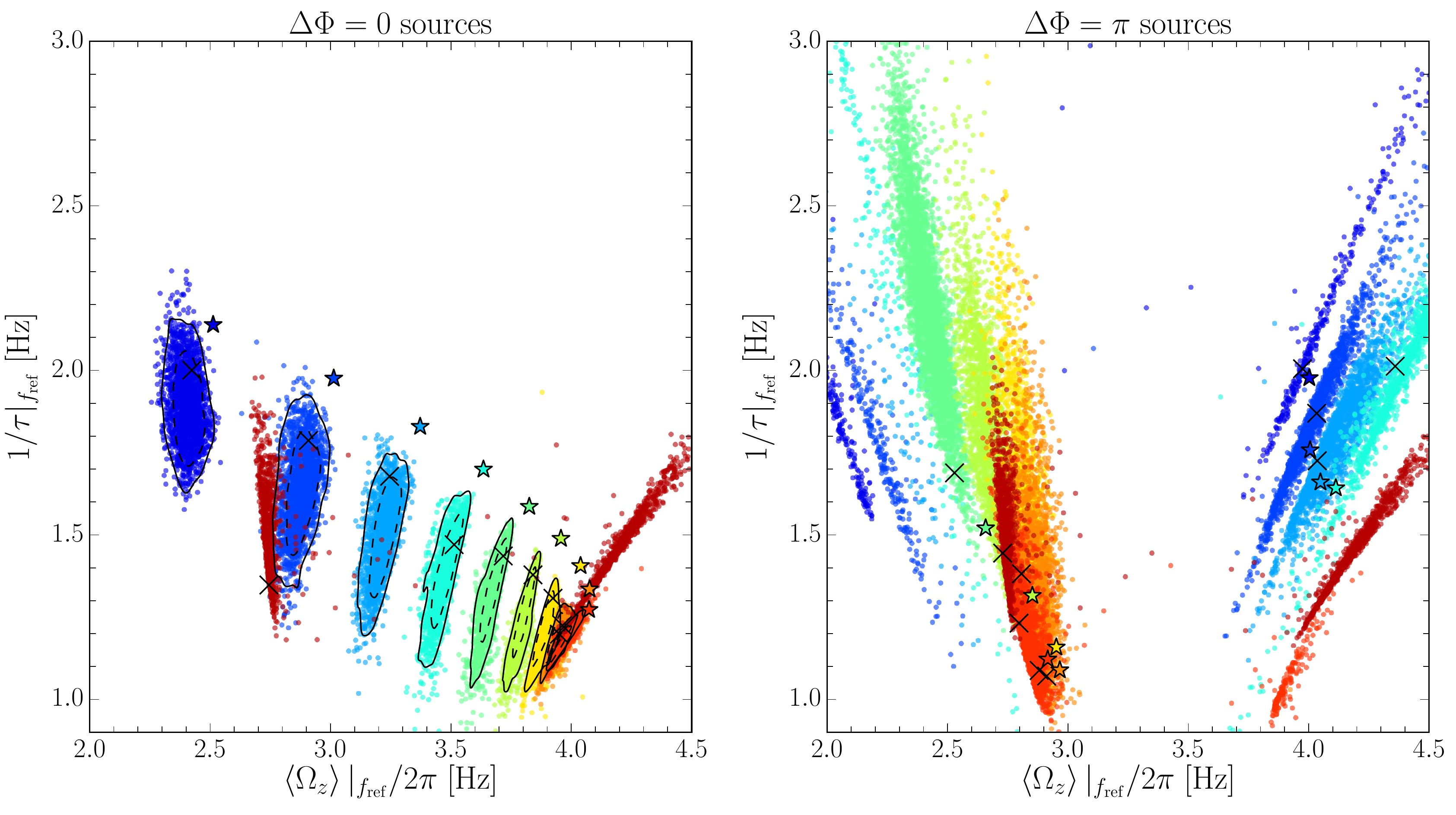}
\end{tabular}
	\caption{\label{fig:OmegaTau}\textbf{Constraining $(\left<\Omega_z\right>,\,\tau)$ at 100~Hz.}  Posterior distributions for the
          $\Delta\Phi=0$ sources (top left) and $\Delta\Phi=\pi$ sources (top right) in the $(\left<\Omega_z\right>/(2\pi),1/\tau)$ plane,
          colored according to simulation $\xi$ value. Values of $\tau$ and $\left<\Omega_z\right>$ are evaluated using the
          binary configuration at the reference frequency $f_{\rm ref}=100$~Hz.           
          Solid (dashed) lines show $95\%$ ($67\%$) confidence intervals for each set of posterior samples from distinct
          simulations. Points are colored according to the $\xi$ values
          of the simulations which are marked by stars.
          Black crosses show, for each set of posterior samples, the maximum likelihood estimate.
          With few exceptions, the posterior associated with each
          $\Delta\Phi=0$ injection is concentrated in a small range of
          $\left<\Omega_z\right>$ and $1/\tau$ near the actual
          value. The results shown here are for $\theta_{JN}=\pi/2$, but similar results hold for
          $\theta_{JN}\ge \pi/4$.
          \emph{Bottom}: Detail of the $\left<\Omega_z\right>/(2\pi),1/\tau$ plane, showing posteriors for simulations with $\xi\geq-0.4$.
	}
\end{figure*}

Figure \ref{fig:OmegaTau} shows the posterior in these ``observable''
parameters: the precession frequencies $\left<\Omega_z\right>$ and
$1/\tau$, evaluated at a GW frequency of $100\unit{Hz}$.
First and foremost, except for outliers
associated with nearly aligned sources and templates, this figure
suggests that, as a first approximation, GW measurements isolate a
candidate source to a relatively narrow region of the
$(\left<\Omega_z\right>,1/\tau)$ plane.  These tight limits are only
possible with sufficient leverage to identify both timescales in the
modulated signal: when the viewing angle is small
($\theta_{JN}\simeq 0$) or precession-induced modulation is small
($\hat{\mathbf L}\cdot \hat{\mathbf J}\simeq 1$, as with
$\Delta\Phi=\pi$; see Figure \ref{fig:Recap}), then observations cannot pin down both timescales.  
Second, because $\left<\Omega_z\right>=\alpha/\tau$ is not
a continuous function of $(J,\xi)$, due to ambiguity in defining the
precession angle $\alpha$ for some degenerate geometries, a posterior
that is connected in the $J\xi$-plane can be disconnected in the
$(\left<\Omega_z\right>,\tau)$ plane.  That said, particularly for the
narrowly confined $\Delta\Phi=0$ sources, the posterior in $\Omega_z$
is connected. %
Third, for spin-orbit resonances, no relative spin precession occurs
-- all spins remain coplanar -- so no modulation on the timescale
$\tau$ exists in the source spin dynamics; hence for all $\theta_{JN}$, no modulations on the timescale $\tau$ can occur
in the source waveform.  We
hypothesize that this lack of modulation, unique to our choice of
sources, contributes to the relative width of the posterior
distributions in $1/\tau$ and $\left<\Omega_z\right>$, though we
cannot verify our hypothesis with this sample.
Extrapolating from our sample, we further anticipate that sources with precession timescales
$\tau$ comparable to or longer than the observationally accessible signal  duration  $ T_{\rm wave}$ also cannot easily be
distinguished from one another.
Based on the limited sample available in our investigation, coordinates that are constant on the precession timescale
like ($J,\xi$) and $(\left<\Omega_z\right>,1/\tau)$ are
particularly valuable tools to extract robust statements about precessing binaries, isolating naturally correlated
and tightly constrained parameters. %
Our methods generalize prior work using seperation-of-timescales  to identify natural parameters and timescales for
precessing black hole-neutron star binaries, and to relate those timescales to GW parameter estimation  \cite{1994PhRvD..49.6274A,gw-astro-SpinAlignedLundgren-FragmentA-Theory,gwastro-mergers-HeeSuk-CompareToPE-Precessing}.
We anticipate these representations will be helpful when mining GW data from precessing BBHs for robust astrophysical
statements.  

\subsection{Biases, the maximum likelihood estimate, and results with noise }

\begin{figure*}
\includegraphics[width=\textwidth]{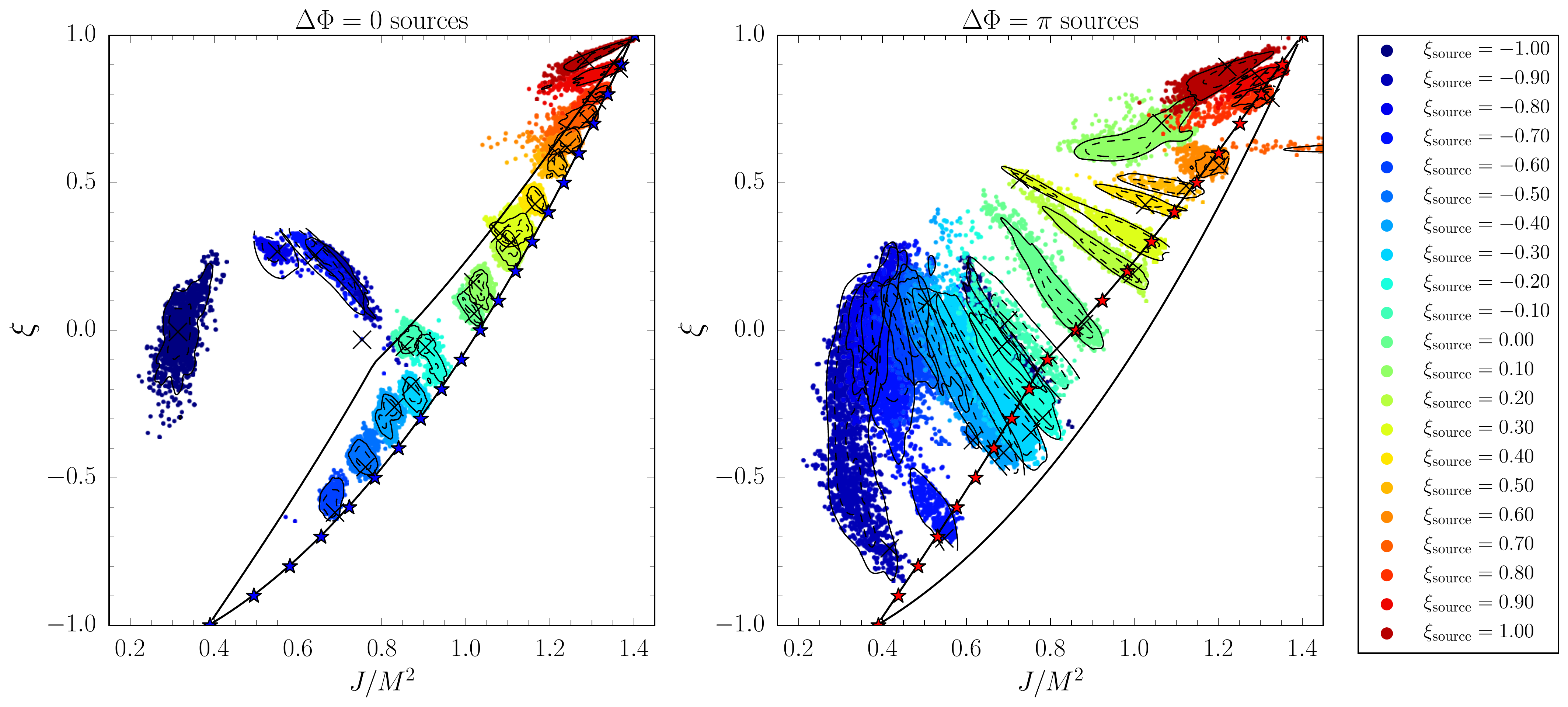}
\caption{\label{fig:ConstrainJXi:Noisy}\textbf{Effect of noise on the posterior.} As Figure \ref{fig:multipanelThetaJN}, this
  figure shows the marginal posterior distribution in $J,\xi$ for sources with $\theta_{JN}=\pi/4$ and SNR=20, but including a random noise realization.
Stars mark the actual values of $J$ and $\xi$ from the simulations. Points and stars are and are colored according
to the simulation's $\xi$ value. 
}
\end{figure*}

Our study used doubly special sources.  On the one hand, resonant
binaries lie in a relatively small corner of the compact binary
parameter space~\cite{2015PhRvL.114h1103K,2015PhRvD..92f4016G}.  On
the other hand, the detector data we used was not a generic noise
realization: for simplicity, we used exactly zero noise.
As a result,  as with posterior distributions of black hole spin and mass ratio for comparable-mass and
highly spinning black holes, we generally cannot expect and do
not observe the posterior confidence intervals in $(J,\xi)$ or $(\Omega_z,\tau)$ to be centered on
the known source parameters: see e.g. Figs.~\ref{fig:ConstrainJXi} and \ref{fig:OmegaTau}. 
Though our posterior distributions had little support for the input
binary's true parameters, our results are nonetheless well-converged
and consistent with a single-best-fitting set of parameters close to
the injected value.  Figures~\ref{fig:ConstrainJXi} and
\ref{fig:OmegaTau} both show a cross $(\times)$ at the location
of the single sample with the maximum likelihood.
Unlike the mean posterior value, our maximum likelihood estimates are generally much closer to the injected value.  
Given how few  posterior samples are available per run, we anticipate that observed differences between the
maximum-likelihood estimates and the injected value are consistent with the limits of our finite sample size.  
Owing to our exceptional inputs and unusual posteriors,
to further demonstrate the stability of our results
we repeated some of our analysis with a random noise realization.
Figure \ref{fig:ConstrainJXi:Noisy} shows our results.  The posterior
distributions remain highly nongaussian, with structures similar to
those observed in the zero-noise study.

\subsection{Morphological classification: Data versus prior}\label{sec:KLdivergence}

\begin{figure*}\centering
\includegraphics[width=\textwidth]{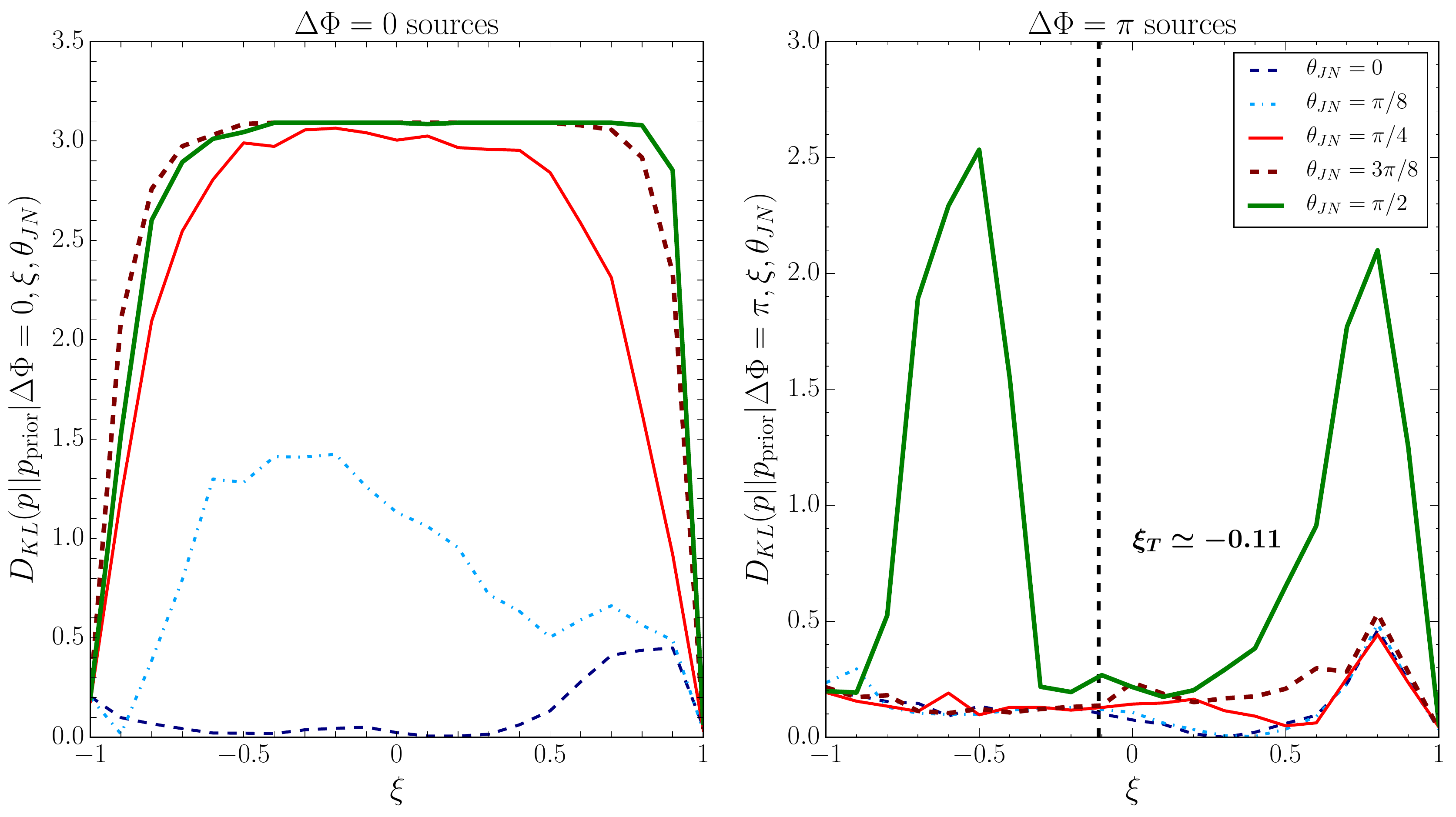}
\includegraphics[width=\textwidth]{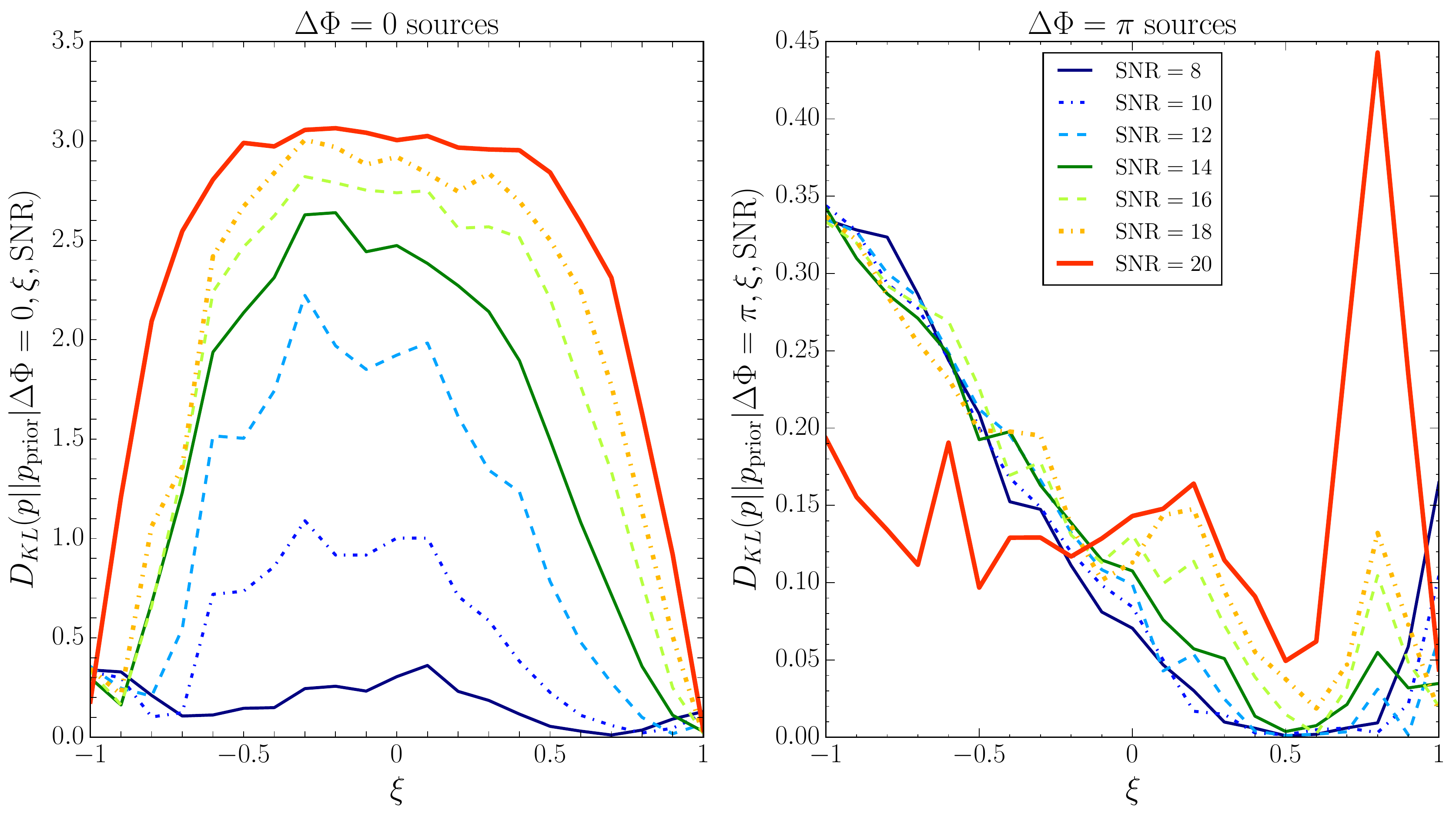}
\caption{\label{fig:KLdivergence}{\bf Information gain.}
  Kullback-Leibler (KL) divergence between the posterior and the prior
  distribution, measured in bits [Eq.~(\ref{eq:DKL})]. \emph{Top:} KL
  divergence as a function of $\xi$ and $\theta_{JN}$ for fixed
  ${\rm SNR}=20$. \emph{Bottom:} KL divergence as a function of $\xi$
  and ${\rm SNR}$ for fixed $\theta_{JN}=\pi/4$.  Structure in these
  plots reflects how the misalignment angles between the various
  angular momenta change versus $\xi$ and how this affects information
  gain: cf. Figures \ref{fig:Recap} and
  \ref{fig:ProbabilityVersusParameters}.
  As described in Appendix \ref{ap:details}, the SNR=20 results (red curve) were produced using a lower starting frequency than
  results for SNR$<20$; this discrepancy is responsible for the slight difference in trend between the bottom left panel's
  result for SNR=20 versus results for SNR$<20$.
  }
\end{figure*}

Unlike conclusions derived about the chirp masses and spins \citeMCMC, for
properties that may be difficult to measure (such as the source
morphology) the prior plays a surprisingly significant role. To
quantify how much information we learn from the data, we compute the
information gain (in bits) as the Kullback-Leibler (KL) divergence
\cite{kullback1951} %
between the morphological classification posterior and prior
distributions:%
\begin{equation}
\label{eq:DKL}
	D_{KL}(p||p_{\rm prior} | \boldsymbol{\theta}) = \sum_{i} p({\cal H}_{i} | \boldsymbol{\theta})
	 \log_{2}\left( \frac{p({\cal H}_{i} | \boldsymbol{\theta})}{p_{\rm prior}({\cal H}_{i} |\boldsymbol{\theta})}\right),
\end{equation}%
where $\boldsymbol{\theta}=\{\Delta\Phi, \xi, \theta_{JN}, \text{SNR}\}$.  The KL
divergence has seen increasing application in physics
\cite{2011RvMP...83..943V,PSconstraints3-MassDistributionMethods-NearbyUniverse,2013MNRAS.435.3521B,2014PhRvD..90b3533S}.
If the posterior resembles the prior ($p\sim p_{\rm prior}$), we have
learned nothing from the data and the KL divergence is nearly $0$.  To
provide a sense of scale, for continuous gaussian distributions with
the same mean but different standard deviations $\sigma\neq \sigma_*$,
the KL divergence between these distributions is
$D_{KL}\simeq [\ln(\sigma/\sigma_*)]^2/\ln 2$; a KL divergence of
order unity therefore implies a difference in mean by one standard
deviation, or a difference in variance by a factor of order $2$.
Figure \ref{fig:KLdivergence} shows this information gain as a function of the source parameters $\Delta\Phi, \xi, \theta_{JN}$.

A comparison with Figure \ref{fig:ProbabilityVersusParameters} reveals
that, despite low posterior probabilities for $\Delta\Phi=\pi$
sources, even in the best case ($\theta_{JN}=\pi/2$) the peak amount
of information we learn from the data is similar for the two source
morphologies.  The posterior probability depends significantly on the
prior.  Specifically, even though the data often strongly favors
${\cal H}_\pi$, the relative rarity of ${\cal H}_\pi$  [Eq. (\ref{eq:prior})] is responsible
for the relatively poor classifications for $\Delta\Phi=\pi$ sources,
compared to $\Delta\Phi=0$ sources.
If we had adopted an astrophysically motivated prior, for example favoring  comparable-mass binaries, we would have
found a far more favorable result for the ability of GW measurements to distinguish between sources in
distinct morphologies. 
We will address the impact of alternative priors in a subsequent study.

\subsection{Systematic uncertainty and conservation of $\xi$}
\label{sec:systematics}
The quantity $\xi$ is known to be conserved at 2PN order by the spin
precession equations when the QM term \cite{2008PhRvD..78d4021R} is
included.  This quantity may not be conserved at higher PN order, and
it is known not to be conserved if the QM term is omitted.
In this paper, we used the default model for black hole spin
precession implemented in \textsc{lalsimulation}, which explicitly
omitted the QM term.  This omission was not intentional -- indeed, all previous analyses also omitted this term \citeMCMC{} --  but it
provides an opportunity to assess the systematic error introduced when
$\xi$ is not exactly conserved on the spin precession timescale, as
could occur at higher PN order.

Using the precession equations adopted in this analysis \cite{1994PhRvD..49.6274A,BCV:PTF},
\begin{align}
\frac{d (\xi L M^2)}{dt}
 = 
 - \frac{3 \mathbf{L}\cdot \mathbf{S}_1\times \mathbf{S}_2 v^7 }{2\eta}
 \mathbf{\hat{L}}\cdot [ \mathbf{S}_1(1+q) - \mathbf{S}_2(1+1/q)]\,.
\end{align}
For configurations at or near a post-Newtonian resonance, $\xi$ will remain nearly constant, because
$\mathbf{L},\mathbf{S}_1$ and $\mathbf{S}_2$ remain nearly coplanar.

\begin{figure*}
\begin{tabular}{c}
\hspace{-1.5cm}\includegraphics[width=0.95\textwidth]{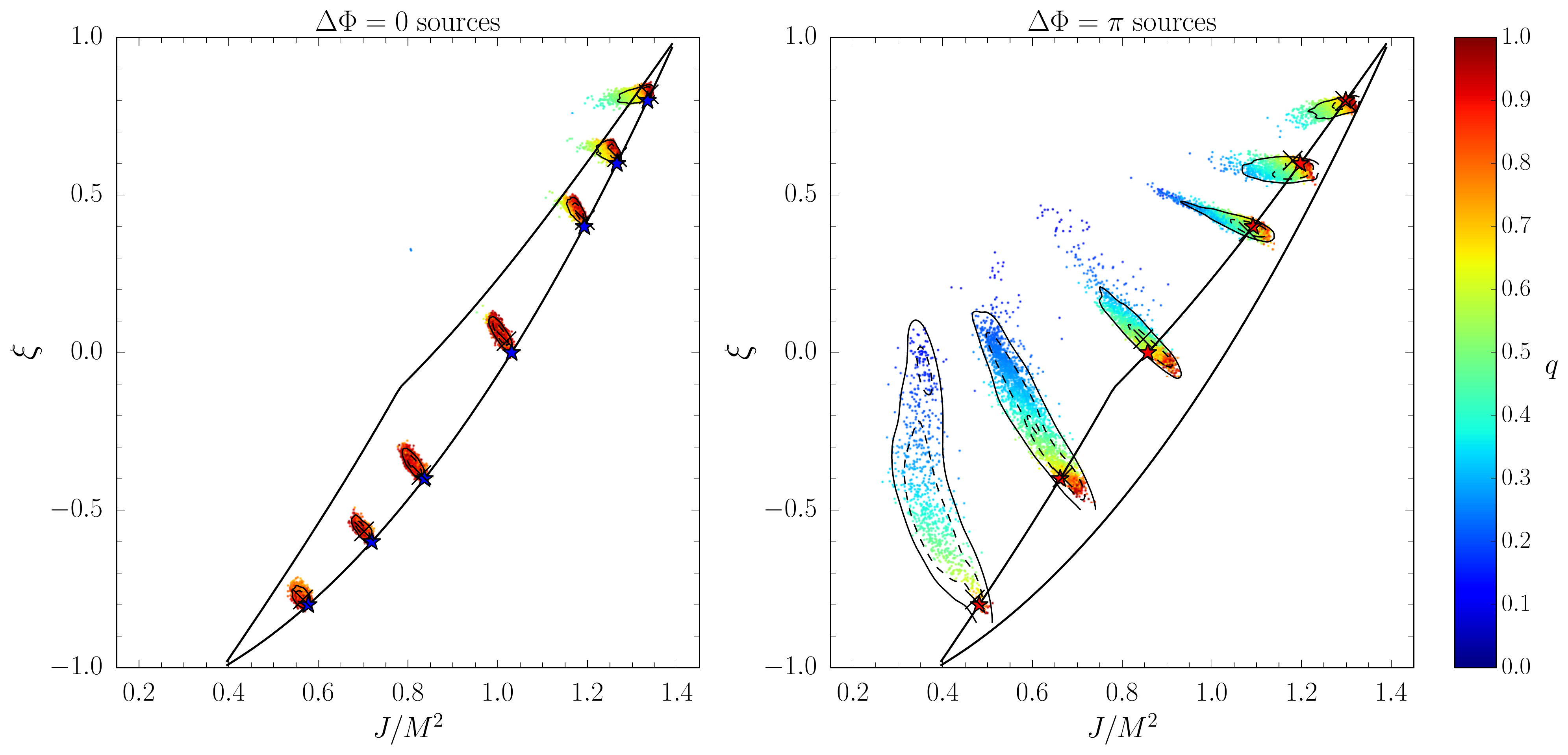}\\
\includegraphics[width=\textwidth]{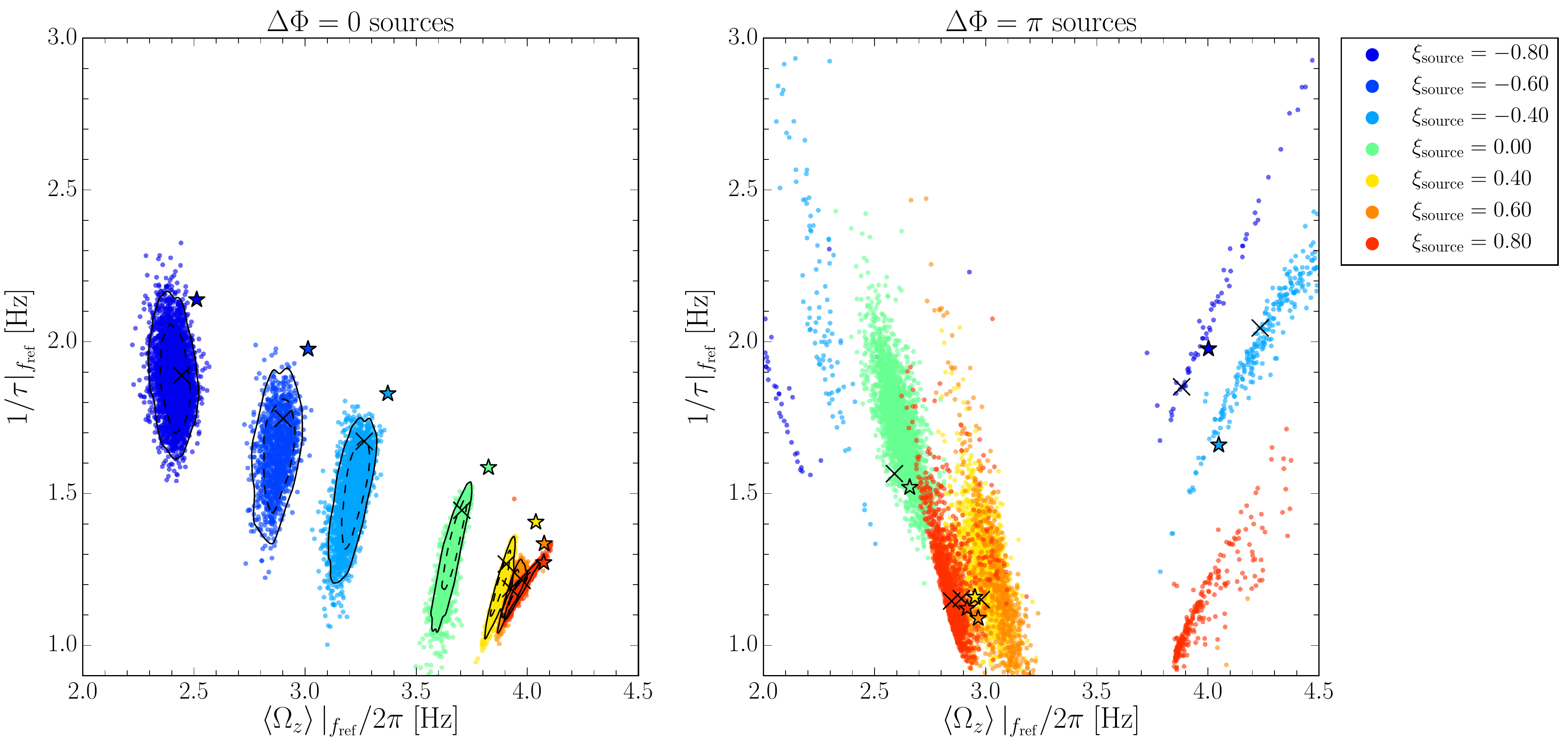}
\end{tabular}
\caption{\label{fig:QMReruns}\textbf{Results using QM-corrected spin precession}: After reproducing selected runs, we
  find similar posteriors in J, $\xi$ (top panel; compare to Figure \ref{fig:ConstrainJXi}) and $\Omega_z,\tau$ (bottom
  panel; compare to Figure \ref{fig:OmegaTau}, which adopts the same color scheme).  
  Consistent with the corresponding figures, the top and bottom panel adopt different color schemes.}
\end{figure*}

For this reason, even though the PN spin precession equations do not
enforce it, for all source binaries $\xi$ is nearly constant
throughout the evolution, rarely varying by as much as $0.05$.
Compared to the typical statistical measurement errors shown in
Figures~\ref{fig:ConstrainJXi}, \ref{fig:multipanelThetaJN} and
\ref{fig:multipanelSNR}, this systematic uncertainty is small for the
signal amplitudes used in this study.  Using a statistically
significant subsample from each posterior distribution, we have also
manually verified that $\xi$ is nearly conserved, occasionally
oscillating but in all cases varying by much less than the statistical
uncertainty in the posterior.  Finally, we have repeated some of the
simulations shown in Figure \ref{fig:ConstrainJXi} with the QM term
included and find quantitatively similar results; see Figure \ref{fig:QMReruns}.  At least for the
resonant sources used in this work, posterior parameter distributions
are not sensitive to the inclusion of the QM term.

\section{\label{sec:Conclusions}Conclusions}

In this work, we use the \textsc{lalinference} parameter estimation code to infer the parameters of precessing BBHs from their GW signal.   
Contrary to some claims that the subdominant spin has little observationally accessible effect \cite{2014PhRvL.112y1101V}, we demonstrate by
concrete example that the distinctive signature of uniquely two-spin physics can be inferred for the special but conservative set of sources we chose,
featuring exactly resonant sources with mass ratio $q=0.8$, $M_{\rm tot}=13.5M_{\odot}$ and maximal spins
Specifically, our analysis demonstrates that GW measurements can infer a subtle feature of the relative
orientation of the two black hole spins: the binary's precessional morphology. 
Each posterior parameter distribution can be subdivided into three
classes, identifying the relative probability that the progenitor
binary had one of three characteristic behaviors (librating about
$\Delta\Phi=0$, librating about $\Delta\Phi=\pi$, or circulating).
Using as sources the most extreme examples of the two librating cases
-- resonant binaries, which exhibit precession on only one of the two
natural timescales -- we confirmed the conclusion
of~\cite{2014PhRvD..89l4025G} that resonant binaries can be
correctly identified with the appropriate morphology.
In other words,  
even though librating spins occupy a small part of parameter space, we show that these measurements reproducibly and
correctly identify the neighborhood associated to each resonance, except for finely tuned highly symmetric
configurations.  

Because of computational cost our study focused on a single set of
masses and spins.  Further investigations with an
astrophysically motivated distribution of masses, spins, and source
distances are needed to assess how often and how reliably morphology
can be classified in practice.
Our promising (if preliminary) results support the hypothesis that
morphological classification could be an astrophysically interesting
and robust diagnostic, as distinct librating morphologies are both
measurable and the natural consequence of distinct features of compact
binary evolution \cite{2013PhRvD..87j4028G}.

Motivated by theoretical studies of precessing binaries \cite{2015PhRvD..92f4016G,2015PhRvL.114h1103K}, we present
our results in a new coordinate system
$(\left<\Omega_z/(2\pi)\right>,1/\tau)$, which better reflects the
physical observables in the GW signal -- precession-induced
modulations in amplitude and phase -- and respects precession physics,
notably separation of timescales.  Based on our limited study of
special systems, this coordinate system seems to better characterize
the dynamical information that GWs provide.

Recently, Ref.~\cite{2014PhRvL.112y1101V} used a small sample to
investigate how well precessing BBH spins could be measured by
Advanced LIGO-scale instruments, using a small number of mass
combinations, spin orientations, and distances ($3\times 2 \times 3$),
but several viewing angles $\theta_{JN}$.  Like previous studies, the
result of~\citet{2014PhRvL.112y1101V} showed that posterior
distributions depend sensitively on viewing angle, being best
constrained when seen edge-on.  Additionally, they investigated how
well subdominant spins could be measured for precessing binaries,
claiming on the basis of a limited sample that the relative angle
$\Delta\Phi$ between the projection of the two spins into the orbital
plane cannot be measured.  Our results provide a concrete
counterexample.  That said, Ref.~\cite{2014PhRvL.112y1101V} used very
different initial configurations with no resonant sources; they
adopted suboptimal diagnostics for the second spin's orientation
(i.e., the time-dependent azimuthal angle $\Delta\Phi$, like Fig.~\ref{fig:ConstrainDeltaPhi}, instead of morphology as
in Fig.~\ref{fig:ProbabilityVersusParameters} or $J,\xi$ point as in Fig.~\ref{fig:ConstrainJXi}); and they had
too few independent initial conditions to robustly explore the full
parameter space.
Specifically, their sample included a handful of sources: three mass pairs, only one of which involved unequal-mass BH-BH binaries ($10M_\odot+5M_\odot$);
fixed BH spin magnitudes; and only two relative
spin orientations, only one of which was not in ${\cal H}_C$ (i.e., not circulating).    For their single noncirculating
source -- a $(10 M_\odot,5 M_\odot)$  pair of rapidly ($\chi_1=\chi_2=0.9$) spinning black holes  with initial spins $\mathbf{S_1}$ oriented by 45$^\circ$ and $\mathbf{S_2}$ oriented
by $135^\circ$ from the line of sight and coplanar with $\mathbf{L}$ -- these authors do note that for $\theta_{JN}\simeq \pi/2$ the posterior
distribution of $\Delta\Phi$ is significantly different from the prior.  
As noted above, further study with an astrophysically motivated distribution of masses, spins, source distances, and orientations is
required to assess how well GWs can identify the source morphology of generic nonresonant sources.

Our analysis employed inspiral-only waveforms which lack the coalescence and ringdown signals present in real BBH merger signals.  Their unphysical termination conditions are known to introduce
convention-dependent artifacts into  parameter estimation, with increasing impact as the total binary mass increases
\cite{2014CQGra..31w5009C,2014CQGra..31o5005M}.   
A detailed discussion of waveform termination conditions is beyond the scope of this paper.  That said, based on the
relatively high termination frequencies shown in Table \ref{tab:fend}, we anticipate
that waveform termination conditions do not dominate the differences we observe.

\begin{acknowledgements}
  D.T. is partially supported by the National Science Foundation
  through awards PHY-1067985, PHY-1404139, PHY-1055103 and PHY-1307020.
  D.T. is grateful for the support and hospitality of V. Kalogera's group and the
  Center for Interdisciplinary Exploration and Research in Astrophysics (CIERA)
  at Northwestern University, where this project was conceived.
  R.O'S. is supported by NSF Award PHY-1505629.
  D.G. is supported by the UK STFC and the Isaac Newton Studentship of the University of
  Cambridge.  E.B.~is supported by NSF CAREER Grant PHY-1055103 and by
  FCT contract IF/00797/2014/CP1214/CT0012 under the IF2014 Programme.
  M.K. is supported by Alfred P. Sloan Foundation grant FG-2015-65299.
  T.B.L.  acknowledges NSF award PHY-1307020.  U.S.~is supported by
  FP7-PEOPLE-2011-CIG Grant No. 293412, FP7-PEOPLE-2011-IRSES Grant
  No.295189, H2020-MSCA-RISE-2015 Grant No.~StronGrHEP-690904,
  H2020 ERC Consolidator Grant Agreement No.~MaGRaTh-646597,
  SDSC and TACC through XSEDE Grant No.~PHY-090003 by the
  NSF, Finis Terrae through Grant No.~ICTS-CESGA-249, STFC Roller
  Grant No. ST/L000636/1 and DiRAC's Cosmos Shared Memory system
  through BIS Grant No.~ST/J005673/1 and STFC Grant Nos.~ST/H008586/1,
  ST/K00333X/1.  Computational resources were provided by the
  Northwestern University Grail cluster (CIERA) through NSF MRI award
  PHY-1126812, by the Atlas cluster at AEI Hannover, supported by
  the Max Planck Institute and by the Nemo at cluster through NSF-092340.
\end{acknowledgements}

\appendix
\section{Coordinate conventions for precessing spins}
\label{ap:coord}
For reference and clarity, in this appendix we summarize the different
coordinate conventions used to describe precessing spins: the
radiation frame, in which angular momenta are parametrized relative to
the line of sight; the ``system frame''
\cite{gwastro-pe-systemframe,gw-astro-PE-lalinference-v1}, in which
angular momenta are parametrized relative to the total angular
momentum $\mathbf{J}$; the spin parameters
$(\theta_{LS_1},\theta_{LS_2},\Delta\Phi)$ frequently used in the
resonant-locking literature, in which the \emph{spin} angular momenta
are parametrized relative to the \emph{orbital} angular momentum
$\mathbf{L}$; and the $(J,\xi,S)$ and ($J,\xi,\varphi')$ coordinate sytems introduced in 
previous work \cite{2015PhRvL.114h1103K,2015PhRvD..92f4016G}. 
Unless
otherwise noted, all parameters are specified at
$f_{\rm ref}=100~\unit{Hz}$.

The radiation frame \cite{BCV:PTF} expresses the orbital and spin
angular momenta in polar coordinates relative to the direction of GW
propagation $-\hat{\mathbf{N}}$, with the zero of azimuthal
angle set by
the plane spanned by $\mathbf{L}, -\hat{\mathbf{N}}$.

In the system frame \cite{gwastro-pe-systemframe}, the angular momenta of the binary are characterized in a hierarchical
fashion.  The angles $\theta_{JN}$ and $\psi_{\rm J}$
are polar coordinates of the total angular momentum $\mathbf{J}$ direction
relative to the direction of propagation $-\hat{\mathbf{N}}$;  one angle $\phi_{JL}$ characterizes the relative orientation of $\mathbf{L}$ on its
precession cone around the total angular momentum, equivalently constraining the direction of the total spin
$\mathbf{S}=\mathbf{S}_1+\mathbf{S}_2$; and three angles $(t_1,t_2,\phi_{12})$ characterize the relative orientation of
the two component spins relative to the orbital angular momentum.   For example, 
$t_1 = \cos^{-1} \hat{\mathbf S}_1 \cdot \hat{\mathbf L} $ and, if $\hat{\mathbf x}$ and $\hat{\mathbf y}$ are unit vectors that form a right-handed
orthonormal frame with $\hat{L}$,
$e^{i \phi_{12}} = {\cal A}\times [(\hat{\mathbf x}+i \hat{\mathbf y})\cdot \hat{\mathbf S}_2]/[(\hat{\mathbf x}+i \hat{\mathbf
    y})\cdot \hat{\mathbf S}_1]$ where ${\cal A}$ is a real amplitude chosen so the norm of the right-hand side is unity.
The resonant-locking literature uses  parameters  $\theta_{1},\theta_{2},\Delta\Phi$ to characterize the relative
orientation of the two spins relative to $\mathbf{L}$. 
These parameters are identical to the tilt and relative phase angles ($t_1=\theta_1 =\theta_{LS_1}$,  $t_2=\theta_2=\theta_{LS_2}$, and
$\Delta\Phi=\phi_{12}$)
used in the system frame currently favored for \textsc{lalinference} \cite{gwastro-pe-systemframe}.
Finally, recent work~\cite{2015PhRvL.114h1103K,2015PhRvD..92f4016G}
provides an explicit solution for the relative two-spin dynamics,
expressing spin evolution in the three-dimensional relative spin space
(e.g., $\theta_{LS_1},\theta_{LS_2},\Delta\Phi$) in yet another
triplet of coordinates $J, \xi, S$ on the three-dimensional space
of all relative spin orientations at fixed $L$, $S_1$, $S_2$, and $q$.  
\citet{2015PhRvD..92f4016G} also introduce an alternative coordinate system $J,\xi,\varphi'$.
In these two systems, the
first coordinate, the magnitude $J$ of the total angular momentum, is
a conserved constant on precession timescales;
the projected effective spin $\xi$ [cf.~Eq.~(\ref{eq:xi})]
is conserved on all timescales by the orbit-averaged 2PN spin precession equations\footnote{Using the complete 2PN spin
  precession equations,
  without orbit averaging,
  $\xi$ varies on the orbital timescale.}; %
the  coordinate
$S$ is the magnitude of the total spin; and the %
coordinate
$\varphi'$ is an angle characterizing the degree to which the plane
spanned by $\mathbf{S}_1, \mathbf{S}_2$ is rotated relative to the
plane spanned by $\mathbf{S},\mathbf{L}$.  Specifically, the angle
$\varphi'$ is defined by decomposing the spins relative to
${\bf S =S_1+S_2}$; a vector perpendicular to ${\bf S,L}$; and the
remaining vector of an orthonormal triad [Fig.~1 in
\cite{2015PhRvD..92f4016G}]:
\begin{eqnarray}
\mathbf{\hat{Z}} \equiv \mathbf{\hat{S} },
\qquad \mathbf{\hat{Y}}  \equiv \frac{\mathbf{L}\times\mathbf{S}}{|\mathbf{L}\times\mathbf{S}|},
\qquad
\mathbf{\hat{X}} = \mathbf{\hat{Y}} \times  \mathbf{\hat{Z}}.
\end{eqnarray}
In terms of these coordinates, we can redundantly express the spin using the spin magnitude $S$ and the relative angle $\varphi'$
between the plane spanned by the two spins and the plane spanned by $\mathbf{S}$ and $\mathbf{L}$:
\begin{subequations}
\label{eq:SpinDecompositionAB}
\begin{align}
\mathbf{S}_1 &= {\cal A} \mathbf{S} + b(\cos \varphi' \mathbf{\hat{X}} + \sin\varphi' \mathbf{\hat{Y}}) \\
\mathbf{S}_2 &= (1-{\cal A}) \mathbf{S} -  b(\cos \varphi' \mathbf{\hat{X}} + \sin\varphi' \mathbf{\hat{Y}})  \\
{\cal A} &= \frac{S^2+S_1^2-S_2^2}{2S^2} \; ; \qquad 1-{\cal A} = \frac{S^2 +S_2^2-S_1^2}{2S^2} \\
{b}^2&=  S_1^2 - {\cal A}^2S^2
\end{align}
\end{subequations}
Either the phase angle $\varphi'$ or the spin $S$ can be eliminated from these expressions; see
\citet{2015PhRvD..92f4016G} for details.

\section{Spin precession formulae}
\label{ap:alpha}

For each value of $m_1,m_2,\chi_1,\chi_2,L,J,\xi$,
Refs.~\cite{2015PhRvL.114h1103K,2015PhRvD..92f4016G} define two
quantities that characterize precession: $\alpha$ and $\tau$.  The
timescale $\tau$ is the duration of one precession cycle of the spins
relative to $\mathbf{L}$, defined by
\begin{eqnarray}
\tau = 2 \int_{S_-}^{S_+} \frac{dS}{|dS/dt|}
\end{eqnarray}
where $S$ is the magnitude of the total spin (cf.~Appendix
\ref{ap:coord}) and where $(dS/dt)^2$ is a polynomial in $S$ with
simple roots at the turning points $S=S_\pm$.  Because $1/|dS/dt|$ has
an integrable singularity at both limits $S_\pm$, $\tau$ is finite and
continuous everywhere, except under certain conditions at the single
point in the $(J, \xi)$ plane %
associated with the larger spin aligned and the smaller
spin antialigned with $\mathbf{L}$
\cite{2015PhRvL.115n1102G}.  %
By contrast, the angle $\alpha$ is the  azimuthal angle accumulated during the precession of  $\mathbf{L}$ around
$\mathbf{J}$.  The azimuthal angle $\phi_{JL}$ of $\mathbf{L}$ around $\mathbf{J}$ is  is therefore not well defined at
times when  $\mathbf{L}$ and $\mathbf{J}$ are  parallel.
As an integral of $d\phi_{JL}/dt$, $\alpha$ inherits this
ambiguity. 
Away from this surface, the function $\alpha$ is well-defined, continuous, and can be evaluated using the definite integral
\begin{eqnarray}
\alpha \equiv  2 \int_{S_-}^{S_+} \Omega_z  \frac{dS}{|dS/dt|}
\end{eqnarray}
where the rational function $\Omega_z$ is given by Eq.~(19) in
\cite{2015PhRvL.114h1103K}.   
This discontinuity in $\alpha$ arises from an integrable singularity
in $\Omega_z$; it implies a discontinuity in $\alpha/\tau$, and it is
responsible for the $V$-shapes seen in Figure \ref{fig:OmegaTau}.

Because the $J\xi$-plane and $\left<\Omega_z\right>=\alpha/\tau$ play a significant role in our presentation, we
provide an explicit formula to determine where these discontinuities occur.   
The conditions on $(J,\xi)$ needed for colinearity follow from Eqs. (6) and (14) in \cite{2015PhRvD..92f4016G}, after substituting
suitable values of $S$.  For clarity, we also derive them from straightforward vector algebra.
Because colinearity of $\mathbf{L}$ and $\mathbf{J}$ implies all angular momenta are coplanar, without loss of
generality we use the polar angles  $\theta_{JS_1}\theta_{JS_2}\in[0,\pi]$ of $\mathbf{S}_{1,2}$ relative to $\mathbf{J}$ to
characterize the two spins.  In these coordinates, colinearity of $\mathbf{L}$ and $\mathbf{J}$ requires
\begin{align}
J &= s L + S_1 \cos \theta_{JS_1} + S_2 \cos \theta_{JS_2} \\
0 &= S_1 \sin \theta_{JS_1} + S_2 \sin \theta_{JS_2} \\ %
\xi &=\frac{s}{M}[m_1 \chi_1\cos \theta_{JS_1} + m_2 \chi_2 \cos \theta_{JS_2}]
\end{align}
where $s=\pm 1$ corresponds to two possible choices (alignment or antialignment) between the orbital and total angular
momentum: $\mathbf{\hat{J}}=s\mathbf{\hat{L}}$

\begin{figure}
\includegraphics[width=\columnwidth]{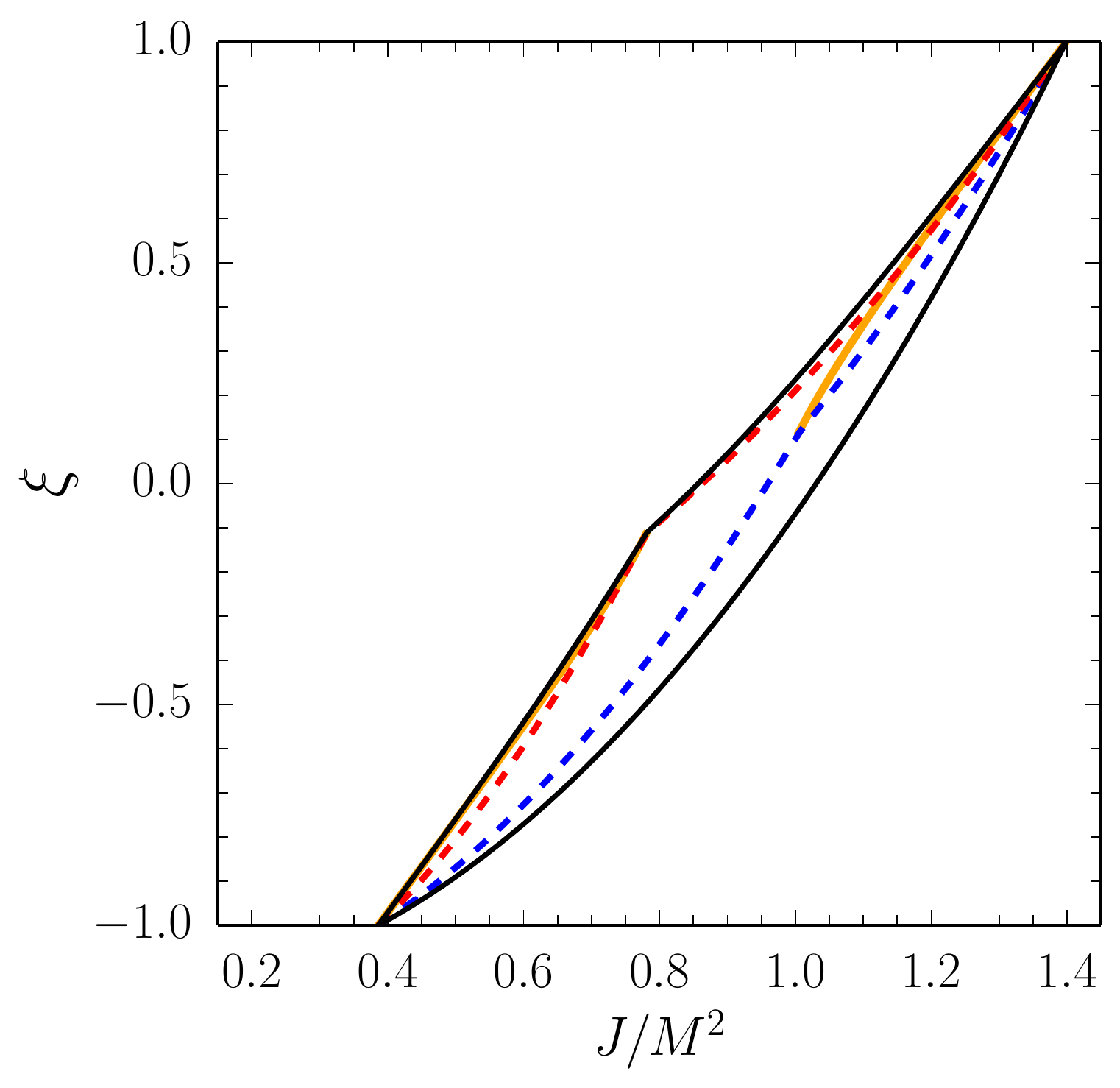}
\caption{\label{fig:AlphaDiscontinuity}\textbf{Alignment between
    $\mathbf{L}$ and $\mathbf{J}$ and discontinuities in $\alpha$ in
    the $J\xi$-plane.} As in Figure
  \ref{fig:BoundaryJXi}, the thick black lines show the allowed region
  in the $J\xi$-plane for our fiducial parameters, and the
  dashed lines show $J_{\rm c0,c\pi}(\xi)$.  The  orange curves show the locus
  of points $J,\xi$ where $\mathbf{L}$ and $\mathbf{J}$ are
  parallel and thus where $\alpha$ is discontinuous, as determined by solutions to Eq.~(\ref{eq:JXi:AlphaDiscontinuityPoint})
  that correspond to physically realizable spin orientations. 
}
\end{figure}

Keeping in mind that $\theta_{JS_1},\theta_{JS_2}\in[0,\pi]$, we find
without loss of generality that these conditions imply one quadratic
equation and one system of two linear equations on the variables
$\cos \theta_{JS_1}, \cos \theta_{JS_2}$:
\begin{align}
S_1^2(1-\cos^2\theta_{JS_1}^2) = S_2^2 (1-\cos \theta_{JS_2}^2) \\
\begin{bmatrix}
J-s L\\ s \xi
\end{bmatrix}
=
 \begin{bmatrix}
m_1^2  & m_2^2 \\
m_1/M & m_2/M
\end{bmatrix}
\begin{bmatrix}
\chi_1 \cos \theta_{JS_1} \\ \chi_2 \cos \theta_{JS_2}
\end{bmatrix}
\end{align}
Solving the linear equations for
$\cos \theta_{JS_1},\cos\theta_{JS_2}$ and then substituting into the
quadratic constraint generally leads to a quadratic form in $(J,\xi)$.
For this problem, however, the coefficient of $\xi^2$ cancels, leaving
a \emph{linear} expression for $\xi$ as a function of $J$. The
solution [compare e.g. to Eq.~(13) in \cite{2015PhRvD..92f4016G}] reads
\begin{eqnarray}
\label{eq:JXi:AlphaDiscontinuityPoint}
\xi &=& s \frac{(J-L s)^2 - \frac{(m_1-m_2)}{m_1+m_2} (S_1^2-S_2^2)} {
2 \eta M^2 (J-L s)
} \nonumber \\
&=& s \frac{
(1+q)^2 (J-L s)^2 - (1-q^2)(S_1^2-S_2^2)
}{
2 q (J - L s) M^2
} .
\end{eqnarray}
Using this solution, we find that the angles
$\theta_{JS_1},\theta_{JS_2}$ vary with $J$ according to
\begin{eqnarray}
 \cos \theta_{JS_1} =  \frac{[(J-L s)^2 + (S_1^2-S_2^2)]}{2 S_1 (J-L s) }, \\
 \cos \theta_{JS_2} =  \frac{[(J-L s)^2 - (S_1^2-S_2^2)]}{2 S_2 (J-L s) }.
\end{eqnarray}
Not all values of $J$ correspond to  realizable angles for $\theta_{JS_1},\theta_{JS_2}$.  
For geometric reasons -- and as can be immediately verified by direct
substitution of appropriate choices for $J$, namely $J=L \pm S_1 \pm S_2$ -- the curve with $s=1$
passes through all four aligned-spin configurations
$|\cos\theta_{JS_1}|=|\cos \theta_{JS_2}|=1$.  Conversely, for $s=1$, a
region containing $J=L$ is always
excluded. %
In the language of~\cite{2015PhRvD..92f4016G}, the root $s=1$
corresponds to $S_{\rm min}=|J-L|$ and the root $s=-1$ corresponds to
$S_{\rm max}=J+L$.
For example, as illustrated by Figure \ref{fig:AlphaDiscontinuity} for our
fiducial parameters, one of these curves (orange in the figure) connects the point $\cos\theta_{JS_1}=\cos\theta_{JS_2}=1$ to the ``kink'' in the $J_{c0}(\xi)$
curve corresponding to $\theta_{JS_1}=\pi,\theta_{JS_2}=0$, the ``up/down'' instability point where $\tau=\infty$
  \cite{2015PhRvL.115n1102G}.  The other curve joins the remaining two spin-aligned configurations and is
nearly equal to the
minimum value $J_{\rm min}(\xi)$ allowed at each $\xi$ over the corresponding range of $\xi$.  
This latter discontinuity is responsible for the significant change in $\Omega_z$ at $\xi=\xic$ seen in Table \ref{tab:fend},
and for the V-shape feature seen in Figure \ref{fig:OmegaTau}.

\begin{table*}
\begin{tabular}{@{\hskip 10pt}c@{\hskip 10pt}c@{\hskip 10pt}c@{\hskip 10pt}|@{\hskip 10pt}c@{\hskip 10pt}c@{\hskip 10pt}|@{\hskip 10pt}c@{\hskip 10pt}c@{\hskip 10pt}c@{\hskip 10pt}|@{\hskip 10pt}c@{\hskip 10pt}c@{\hskip 10pt}}
$\Delta\Phi$ & $\xi$ & $J/M^2$ & $f_{\rm end} (\unit{Hz})$ & ${\Delta \phi_{JL}}/2\pi$ & $\theta_{JL}$ & $\theta_{LS_1}$ & $\theta_{LS_2}$ 
 & ${\left<\Omega_z\right>}/{2\pi}$[Hz] & $1/\tau$[Hz] \\\hline
 0  &        -1.0  &        0.39  &         281 &       3.50  &        0.00  &        $\pi$  &        $\pi$  &         -  &         -  \\ 
 0  &        -0.9  &        0.49  &         303 &       3.98  &        0.46  &        2.63  &        2.79  &        2.13  &        2.16  \\ 
 0  &        -0.8  &        0.58  &         321 &       4.47  &        0.55  &        2.41  &        2.63  &        2.46  &        2.07  \\ 
 0  &        -0.7  &        0.66  &         345 &       4.83  &        0.58  &        2.23  &        2.51  &        2.73  &        1.99  \\ 
 0  &        -0.6  &        0.72  &         375 &       5.20  &        0.59  &        2.09  &        2.39  &        2.95  &        1.92  \\ 
 0  &        -0.5  &        0.78  &         403 &       5.59  &        0.59  &        1.95  &        2.29  &        3.14  &        1.84  \\ 
 0  &        -0.4  &        0.84  &         438 &       5.94  &        0.58  &        1.83  &        2.19  &        3.30  &        1.77  \\ 
 0  &        -0.3  &        0.89  &         483 &       6.25  &        0.56  &        1.71  &        2.09  &        3.44  &        1.71  \\ 
 0  &        -0.2  &        0.94  &         533 &       6.58  &        0.55  &        1.60  &        1.99  &        3.55  &        1.65  \\ 
 0  &        -0.1  &        0.99  &         608 &       6.88  &        0.52  &        1.50  &        1.90  &        3.65  &        1.59  \\ 
 0  &         0.0  &        1.03  &         687 &       7.18  &        0.50  &        1.39  &        1.80  &        3.74  &        1.54  \\ 
 0  &         0.1  &        1.08  &         763 &       7.47  &        0.47  &        1.29  &        1.69  &        3.81  &        1.49  \\ 
 0  &         0.2  &        1.12  &         812 &       7.73  &        0.45  &        1.19  &        1.59  &        3.87  &        1.44  \\ 
 0  &         0.3  &        1.16  &         812 &       7.99  &        0.42  &        1.09  &        1.48  &        3.91  &        1.40  \\ 
 0  &         0.4  &        1.20  &         834 &       8.22  &        0.39  &        0.99  &        1.36  &        3.95  &        1.36  \\ 
 0  &         0.5  &        1.23  &         823 &       8.48  &        0.35  &        0.88  &        1.23  &        3.97  &        1.33  \\ 
 0  &         0.6  &        1.27  &         825 &       8.67  &        0.31  &        0.77  &        1.10  &        3.99  &        1.30  \\ 
 0  &         0.7  &        1.30  &         831 &       8.90  &        0.27  &        0.66  &        0.95  &        3.99  &        1.26  \\ 
 0  &         0.8  &        1.34  &         865 &       9.11  &        0.22  &        0.53  &        0.77  &        3.99  &        1.23  \\ 
 0  &         0.9  &        1.37  &         857 &       9.29  &        0.16  &        0.37  &        0.54  &        3.97&        1.21  \\ 
 0  &         1.0  &        1.40  &         843 &       9.47  &        0.00  &        0.00  &        0.00  &          -
 &         -  \\  \hline
$\pi$ &        -1.0  &        0.39  &         281 &       6.29  &        0.00  &        $\pi$  &        $\pi$  &         -  &         -  \\ 
$\pi$ &        -0.9  &        0.44  &         299 &       6.44  &        0.04  &        2.84  &        2.55  &       -&         -  \\ 
$\pi$ &        -0.8  &        0.48  &         317 &       6.62  &        0.06  &        2.74  &        2.27  &        3.92  &        1.92  \\ 
$\pi$ &        -0.7  &        0.53  &         336 &       6.83  &        0.08  &        2.69  &        2.04  &        3.91  &        1.80  \\ 
$\pi$ &        -0.6  &        0.58  &         359 &       7.05  &        0.09  &        2.67  &        1.81  &        3.92  &        1.71  \\ 
$\pi$ &        -0.5  &        0.62  &         384 &       7.30  &        0.10  &        2.68  &        1.58  &        3.94  &        1.65  \\ 
$\pi$ &        -0.4  &        0.67  &         413 &       7.57  &        0.10  &        2.72  &        1.33  &        3.96  &        1.61  \\ 
$\pi$ &        -0.3  &        0.71  &         444 &       7.92  &        0.09  &        2.78  &        1.05  &        3.99  &        1.60  \\ 
$\pi$ &        -0.2  &        0.75  &         481 &       8.29  &        0.07  &        2.89  &        0.71  &        4.02  &        1.60  \\ 
$\pi$ &        -0.1  &        0.79  &         518 &       4.51  &        0.06  &        2.92  &        0.10  &        2.47  &        1.59  \\ 
$\pi$ &        -0.0  &        0.86  &         569 &       4.80  &        0.16  &        2.43  &        0.32  &        2.60  &        1.48  \\ 
$\pi$ &         0.1  &        0.92  &         636 &       5.12  &        0.19  &        2.14  &        0.44  &        2.71  &        1.38  \\ 
$\pi$ &         0.2  &        0.98  &         712 &       5.44  &        0.20  &        1.91  &        0.53  &        2.79  &        1.28  \\ 
$\pi$ &         0.3  &        1.04  &         755 &       5.72  &        0.19  &        1.69  &        0.60  &        2.85  &        1.20  \\ 
$\pi$ &         0.4  &        1.10  &         799 &       5.96  &        0.17  &        1.49  &        0.65  &        2.89  &        1.13  \\ 
$\pi$ &         0.5  &        1.15  &         819 &       6.23  &        0.15  &        1.30  &        0.67  &        2.91  &        1.08  \\ 
$\pi$ &         0.6  &        1.20  &         834 &       6.46  &        0.13  &        1.11  &        0.66  &        2.90  &        1.06  \\ 
$\pi$ &         0.7  &        1.25  &         850 &       6.65  &        0.10  &        0.92  &        0.62  &        2.89  &        1.07  \\ 
$\pi$ &         0.8  &        1.30  &         864 &       6.80  &        0.08  &        0.72  &        0.53  &        2.86  &        1.09  \\ 
$\pi$ &         0.9  &        1.35  &         879 &       6.91  &        0.05  &        0.49  &        0.39  &        2.82  &        1.13  \\ 
$\pi$ &         1.0  &        1.40  &         843 &       6.98  &        0.00  &        0.00  &        0.00  &         -  &         -  \\ 
\end{tabular}
\caption{\label{tab:fend}\textbf{Derived source properties.} For each  dynamically distinct binary used as a source of
  GWs, this
  table reports on properties of the dynamics, including: $f_{\rm end}$, the termination frequency of the GW signal; $\Delta\phi_{JL}$,
  the number of precession cycles of $\mathbf{L}$ around $\mathbf{J}$ between the starting and ending frequencies; 
  $\theta_{JL},\theta_{LS_1},\theta_{LS_2}$, characterizing the relative orientations of the angular momenta
  $\mathbf{J,L,S_1,S_2}$;
  $J$, the magnitude of the total angular momentum; and $\left<\Omega_z\right>/(2\pi)$ and $1/\tau$, the characteristic
  precession frequencies.  The last five columns are evaluated at the reference frequency, $100\unit{Hz}$.  
  Discontinuities in $\left<\Omega_z\right>$ occur as described in Appendix \ref{ap:alpha}.
}
\end{table*}

\section{Technical details and caveats}
\label{ap:details}

Small technical details about waveform generation and sampling can have a
dramatic impact on final results.  In this appendix, we provide more exhaustive details about the specific calculations
we perform.

All signals are sampled with an interval $\Delta t  = 1/2048 \unit{s}$, corresponding to a Nyquist frequency
$f_{\rm Nyq}=1024\unit{Hz}$.  Our signal and templates start at
$2f_{\rm orb} =10\unit{Hz}$.  [For sources with SNR $<20$, we used $2 f_{\rm orb}=20 \unit{Hz}$.]
Each binary is  evolved until terminated at the minimum stable circular orbit or when
the orbital frequency begins to decrease with time, whichever comes first.   For the masses and spins studied here, this termination frequency is
significantly smaller  than the Nyquist frequency, but significantly higher than the frequency at which most power
accumulates; see, e.g., Fig 4 in \cite{2014PhRvD..89l4025G}.

Our signals were analyzed in a $128\unit{s}$ data window, far longer than the signal.  [For sources with  SNR $<20$, we
  used $16\unit{s}$.] As described elsewhere, rather
than use the full 15-dimensional likelihood, we
explicitly marginalized it over time and orbital phase at each step in the MCMC
\cite{gw-astro-PE-lalinference-v1}. Using selected examples, we have
confirmed that our results are unchanged if this
marginalization is not used.  
As seen in Table~\ref{tab:fend}, particularly for
$\xi \simeq -1$, some of our signals were short, being
terminated at comparatively small frequencies due to the breakdown of
the post-Newtonian expansion we employ to generate them.  That said,
not only is relatively little signal power associated with
$f\gtrsim 300 \unit{Hz}$, but also no structure in the posterior
correlates tightly with the termination frequencies listed in Table
\ref{tab:fend}.  A detailed study of the impact of termination
conditions on precessing parameter estimation is beyond the scope of
our work.

With the exception of Figure~\ref{fig:ConstrainJXi:Noisy}, in the text
we use an exactly zero-noise realization.  The likelihood of a given
set of GW detector data depends on the detector noise in two ways
\cite{gw-astro-PE-lalinference-v1,gwastro-mergers-HeeSuk-CompareToPE-Aligned}:
through the specific detector data being analyzed, and through the
probability of any given noise realization.  Each instrument's noise
$n(t)$ is assumed to be stationary and gaussian Markov
process, %
characterized by a noise power spectrum
$\left<\tilde{n}(\omega')^*\tilde{n}(\omega)\right> =
S_h(\omega)\delta(\omega-\omega')/2$;
using the noise power spectrum, we can evaluate the probability of any
noise realization $n(t)$.  To synthesize a unique data set to be
analyzed for each set of intrinsic parameters $\lambda$, we assume
that the data $d(t)$ in each instrument -- generally containing both
signal and noise, or of the form $d(t)=h(t|\lambda)+n(t)$ for each
instrument --- is given exactly by
$d(t)=h(t)=F_+h_+(t|\lambda)+F_\times h_\times(t|\lambda)$, where
$F_{+,\times}$ are the detector response functions and
$h_{+,\times}(t|\lambda)$ are the two linear polarizations of of the
GW.  Just like evaluating a normal distribution at the mean, this
arbitrary choice eliminates ambiguity in subsequent Bayesian
calculations of the posterior distribution.

\bibliography{paper}

\end{document}